# Colossal dielectric response of $Hf_xZr_{1-x}O_2$ nanoparticles

Oleksandr S. Pylypchuk[1], Victor V. Vainberg[1*], Vladimir N. Poroshin[1], Oksana V. Leshchenko[2], Victor N. Pavlikov[2], Irina V. Kondakova[2], Serhii E. Ivanchenko[2], Lesya P. Yurchenko[2], Lesya Demchenko[3,4], Anna O. Diachenko[5], Myroslav V. Karpets[2,4], Eugene A. Eliseev[2†], and Anna N. Morozovska[1‡]

[1] Institute of Physics, National Academy of Sciences of Ukraine, 46, Nauky Avenue, 03028 Kyiv, Ukraine

[2] Frantsevich Institute for Problems in Materials Science, National Academy of Sciences of Ukraine, 3, str. Omeliana Pritsaka, 03142 Kyiv, Ukraine

[3] Stockholm University, Department of Chemistry, Sweden

[4] Ye. O. Paton Institute of Materials Science and Welding, National Technical University of Ukraine "Igor Sikorsky Kyiv Polytechnic Institute", 37, Beresteisky Avenue, Kyiv, Ukraine, 03056

[5] Oles Honchar Dnipro National University, 72, Nauky Avenue, 49000 Dnipro, Ukraine

**Abstract**

We observed a colossal dielectric response of small (5 – 10 nm) oxygen-deficient $Hf_xZr_{1-x}O_2$ nanoparticles (x = 1 – 0.4), prepared by the solid-state organonitrate synthesis. The effective dielectric permittivity of the pressed $Hf_xZr_{1-x}O_2$ nanopowders has a pronounced maximum at 38 – 88ºC, which shape can be fitted by the Curie-Weiss type dependence modified for the diffuse ferroelectric-paraelectric phase transition. The maximal value of the dielectric permittivity increases from $1.5 \cdot 10^3$ (for x = 1) to $1.5 \cdot 10^5$ (for x = 0.4) at low frequencies (~4 Hz); being much smaller, namely changing from 7 (for x = 1) to 20 (for x = 0.4) at high frequencies (~500 kHz). The frequency dispersion of the dielectric permittivity maximum position is almost absent, meanwhile the shape and width of the maximum changes in a complex way with increase in frequency. The temperature dependencies of the dielectric permittivity and resistivity are almost mirror-like turned over in respect to each other, which means that all their features, such as position and shape of maxima, plateau, minima and inflexions, almost coincide after the mirror reflection in respect to the temperature axis. These correlations of resistivity and dielectric permittivity are well-described in the Heywang barrier model applied together with the variable range hopping conduction model in semiconducting ferroelectrics. The ferroelectric-like behavior of the small oxygen-deficient $Hf_xZr_{1-x}O_2$ nanoparticles is expected from the Landau-Ginzburg-Devonshire approach and density functional theory calculations. Obtained results may be useful for developing silicon-compatible functional nanomaterials based on $Hf_xZr_{1-x}O_2$ nanoparticles.

---

* Corresponding author: viktor.vainberg@gmail.com
† Corresponding author: eugene.a.eliseev@gmail.com
‡ Corresponding author: anna.n.morozovska@gmail.com

## I. INTRODUCTION

Nowadays thin films [1], multilayers [2] and superlattices [3] of hafnia ($HfO_2$) and zirconia ($ZrO_2$), and their solid solutions $Hf_xZr_{1-x}O_2$ (0<x<1) become indispensable Si-compatible ferroelectric materials [4] used in the advanced electronic memories [5] and logic devices, including steep-slope or/and negative capacitance field-effect transistors with the hafnia-zirconia gate oxides [6, 7]. In particular, Cheema et al. [8] demonstrated the pronounced ferroelectric properties of ultrathin $Hf_{0.8}Zr_{0.2}O_2$ films synthesized by atomic layer deposition on Si-substrate buffered with a $SiO_2$ layer. Also, the thickness-induced antiferroelectric-to-ferroelectric phase transition in thin $ZrO_2$ films on Si substrate was revealed experimentally and the ferroelectricity in ultrathin (5 Å thick) $ZrO_2$ film was observed [9]. Later Cheema et al. [10, 11] studied experimentally polar properties and charge storage in ferroelectric-antiferroelectric $HfO_2$/$ZrO_2$ multilayers and superlattices deposited on Si substrate and demonstrated the capacitance enhancement via the negative capacitance effect.

However, the underlying physical mechanisms of the ferroelectric, ferrielectric and/or antiferroelectric properties emergence in the nanoscale $Hf_xZr_{1-x}O_2$ are still under debate [12, 13, 14] and most probably have an extrinsic nature [15] related with size effects. In particular, the anomalous size dependence of ferroelectricity in $Hf_xZr_{1-x}O_2$, which contradicts the conventional ferroelectric behavior, is conditioned by their fluorite-type structure that enables surface energy driven phase transition at nanoscale, unlike e.g., perovskite ferroelectrics, where the depolarization effects suppress or reduce ferroelectricity with decrease in their size [1-5].

Due to the size effects, the vast majority of ferroelectric crystalline materials exhibit ferroelectric properties in the bulk and become non-polar (paraelectric or dielectric) when their actual size decreases below the critical size: the co-called size-induced transition to the non-polar phase happens [16]. The hafnia-zirconia is a very rare exception, because bulk $HfO_2$ and $ZrO_2$ single-crystals are dielectrics from low to high temperatures (up to 1200 K) and pressures (up to 15 GPa) [17, 18]. The bistable spontaneous polarization (or antiferroelectric long-range order), domain structure and ferroelectric (or antiferroelectric) hysteresis can appear in the $Hf_xZr_{1-x}O_2$ (0≤x≤1) thin films only with a decrease in their thickness to nanoscale [19, 20]. At that the ferroelectric state stabilizes in a rather narrow thickness range, from 5 to 30 nm [21]. The origin of the thickness-induced ferroelectricity is related to the film transition from the non-polar monoclinic m-phase (space group *P2₁/c*) to the polar orthorhombic o-phase (space group *Pca2₁*). The ferroelectric o-phase is metastable in comparison to the bulk m-phase.

The ferroelectric-antiferroelectric properties of $Hf_xZr_{1-x}O_2$ thin films are very sensitive to the content "x" of Hf atoms [22], interface effects [23], synthesis methods and doping [24, 25]. Recent experimental studies [13, 26, 27] and theoretical calculations [28, 29] revealed a significant role of the oxygen vacancies [30], surface effects and grain size [31] in the stabilization of the ferroelectric o-phase in $Hf_xZr_{1-x}O_2$ thin films. The density functional theory (DFT) calculations [32, 33] confirmed that the m-phase is the ground state in the bulk $Hf_xZr_{1-x}O_2$, however the energy difference between the ferroelectric o-phase and non-polar m-phase can be rather small (e.g., ~20 meV/f.u.[33]). Latest DFT calculations reveal that indirect switching paths of the spontaneous polarization are energetically preferable in $HfO_2$ due to competing m- and o-phases [34, 35]. Phenomenological Landau-Ginzburg-Devonshire (LGD) theory predicted [36] that poly-domain or single-domain ferro-ionic states can be stable in spherical $Hf_xZr_{1-x}O_2$ nanoparticles of sizes 5 – 30 nm covered with surface ions and/or oxygen vacancies, when their concentration exceeds $10^{17} - 10^{19}$ cm$^{-2}$.

To the best of our knowledge, any direct experimental observation of the ferroelectric (or ferroelectric-like) properties of $Hf_xZr_{1-x}O_2$ nanoparticles are still absent. A few observations of the o-phases mixture (*Pca*$2_1$, *Pbca* and *Pbcm*) in the small size (3 – 30 nm) $Hf_xZr_{1-x}O_{2-y}$ nanoparticles were obtained by the X-ray diffraction (XRD) analysis [37, 38, 39]. However, these experimental observations are not convincing, because only the o-phase *Pca2$_1$* is ferroelectric; the o-phases *Pbca* and *Pbcm* are neither polar nor ferroelectric-like. Little more convincing (but indirect) evidence is the charge accumulation effect observed in $Hf_{0.5}Zr_{0.5}O_2$ nanopowders, sintered by the auto-combustion sol-gel method, which were related with the appearance of the ferroelectric-like polar regions in the nanoparticles.

In this work we study a dielectric response of small (5 – 10 nm) oxygen-deficient $Hf_xZr_{1-x}O_2$ nanoparticles (x = 1, 0.6, 0.5 and 0.4) prepared by the solid-state organonitrate synthesis, where the domination of the o-phase was proved by the XRD. We reveal that the effective dielectric permittivity of the pressed $Hf_xZr_{1-x}O_2$ nanopowders has a pronounced maximum at 38 – 88°C, which shape can be fitted by the Curie-Weiss type law modified for the diffuse ferroelectric-paraelectric phase transition. The temperature dependencies of the dielectric permittivity and resistivity are almost mirror-like turned over in respect to each other; in the case of resistivity being in quite good agreement with the Heywang [40] barrier model applied together with the variable range hopping conduction model in semiconducting ferroelectrics. The observed behavior of the dielectric permittivity can be explained by the LGD approach and DFT calculations, which reveal that small $Hf_xZr_{1-x}O_2$ nanoparticles can become ferroelectric in the presence of oxygen vacancies.

## 2. SAMPLES PREPARATION AND CHARACTERIZATION
### A. Preparation of the oxygen-deficient $Hf_xZr_{1-x}O_2$ nanoparticles and their characterization

Using the solid-state organonitrate synthesis, we prepared several sample groups of oxygen-deficient $Hf_xZr_{1-x}O_2$ (x=1, 0.4., 0.5 and 0.6) nanopowders. The preparation details of the samples are listed in **Table I**. For this purpose, the mixtures of zirconium and hafnium nitrate salts, $ZrO(NO_3)_2$ $2H_2O$ and $Hf(NO_3)_2$ $2H_2O$, were dissolved in distilled water to a concentration of 5 – 10 % or less. The dextrin, glycine, polyvinyl alcohol, cellulose, etc. were used as organic additives. Based on the combination of properties and synthesis parameters, the fibrous cellulose was selected in the form of ashless filters of various densities and cotton cosmetic discs, which were impregnated with solutions of zirconium and hafnium nitrates and dried at 90°C to remove water. The compositions obtained in this way were subjected to heat treatment in a $CO + CO_2$ atmosphere at 500-600°C for several hours. The black color of the pyrolysis products is due to the presence of the (20 – 50) wt. % ultrafine carbon, which was removed by the short-term oxidation at 600°C for 5 minutes.

The XRD analysis was performed using the XRD-6000 diffractometer with Cu-Kα1 emission (2θ = 15 - 70°); and the database of the International Committee for Powder Diffraction Standards (JCPDS PDF-2) was used for identification of the crystallographic phases of the $Hf_xZr_{1-x}O_2$ nanoparticles. In the result, we confirmed a coexistence of the m-phase (space group P21/c, the content varies from 13 to 5 wt/ %) and inseparable o-phases (space groups Pbca, Pbcm and ferroelectric Pca21, the content varies from 87 to 96 wt. %) inside the nanoparticles (see **Table I**). Note that the o-phase content is maximal for the $Hf_{0.5}Zr_{0.5}O_2$ sample annealed at 600°C for 2 hours in $CO+CO_2$, and minimal for the $HfO_2$ sample annealed at 500°C for 6 hours in $CO+CO_2$. The size of the coherent scattering regions (CSR) varies from 80 to 100 Å for the o-phases, and from 120 to 170 Å for the m-phase. Corresponding diffractograms are presented in **Fig. S1** in **Appendix S1** [41]**.**

**Table I.** Characteristics of the $Hf_xZr_{1-x}O_2$ nanopowders

| Sample name and chemical composition | Characteristics of the samples from XRD | | | | Preparation details |
|---|---|---|---|---|---|
| | Lattice parameters (Å)* | | phase (wt. %), ** CSR size (Å) | | |
| | o-phases | m-phase | o-phases | m-phase | |
| HZO1 $HfO_2$ | $a$ = 10.118 $b$ = 5.202 $c$ = 5.122 | $a$ = 5.125 $b$ = 5.158 $c$ = 5.305 | 87 % 80 Å | 13 % 120 Å | decomposition of nitrates at 500°C, 6 hours in CO + $CO_2$ |
| HZO2 $Hf_{0.6}Zr_{0.4}O_2$ | $a$ = 10.108 $b$ = 5.128 $c$ = 5.160 | $a$ = 5.148 $b$ = 5.167 $c$ = 5.331 | 95 % 85 Å | 5 % 150 Å | decomposition of nitrates at 600°C, 2 hours in CO + $CO_2$ |
| HZO3 $Hf_{0.5}Zr_{0.5}O_2$ | $a$ = 10.062 $b$ = 5.128 $c$ = 5.160 | $a$ = 5.146 $b$ = 5.163 $c$ = 5.328 | 96 % 100 Å | 4 % 140 Å | decomposition of nitrates at 600°C, 2 hours in CO + $CO_2$ |
| HZO4 $Hf_{0.4}Zr_{0.6}O_2$ | $a$ = 10.109 $b$ = 5.128 $c$ = 5.160 | $a$ = 5.130 $b$ = 5.208 $c$ = 5.307 | 89 % 90 Å | 11 % 170 Å | decomposition of nitrates at 600°C, 2 hours in CO + $CO_2$ |

*The monoclinic angle $\beta = 99.2$ degrees for the m-phase. This angle was fixed for all samples during the refinement process.

**As it was shown in earlier studies of $Hf_{0.5}Zr_{0.5}O_2$ nanoparticles performed by Fujimoto et al. [37], the decomposition of the XRD spectra can be done using the "m + o" basis and the "m + t" basis, where "t" means the tetragonal phase. It is impossible to distinguish the t-phase and O-phase in the $Hf_xZr_{1-x}O_2$ nanopowders because corresponding XRD peaks are diffused in small nanoparticles. In agreement with Ref. [37], the decomposition of the XRD spectra using the full "m + t + o" basis performed in Ref. [39], appeared doubtful due to the proximity of the o- and t-phases spectral lines. In this work we start fitting the XRD spectrum using "m + o" basic and reach a good convergency with different lattice constants $a$, $b$ and $c$, listed in **Table I**. The condition $a = b$ should be valid in the t-phase. When one start from the "m + t + o" basic in the considered case of the oxygen-deficient $Hf_xZr_{1-x}O_2$ nanopowders, the axes $a$ and $b$ do not become equal at the end of the fitting time. We have tested this result for all samples, which confirmed that we have the "o + m" case, probably due to the large amount of oxygen vacancies stabilizing the o-phases.

The higher content of the monoclinic phase in the $Hf_{0.4}Zr_{0.6}O_2$ nanopowders compared to the $Hf_{0.6}Zr_{0.4}O_2$ nanopowders can be related with the different stability of the m-phase and o-phases in the presence of oxygen vacancies, which concentration can depend on the Hf content "x". As a matter of fact, the concentration of oxygen vacancies, which can stabilize the polar o-phase [29, 42, 43, 44], indeed depends on the content "x" in the studied samples [45]. This assumption is confirmed by the x-dependence of the defect-related luminescence in Raman spectra, and in the x-dependence of the saturation magnetization, induced by paramagnetic oxygen vacancies [45]. The increase of the m-phase fraction in the $Hf_{0.4}Zr_{0.6}O_2$ nanopowders correlates with decrease in the vacancy concentration in the particles [45]. Namely, the defect-related luminescence and the magnetization is smaller in the $Hf_{0.4}Zr_{0.6}O_2$ nanopowder than that of $Hf_{0.6}Zr_{0.4}O_2$ nanopowder (see Figs.2(a), Table I and Fig. 3(a) in Ref.[45]).

A clear distinction between the polar Pca2₁ and the centrosymmetric Pbca/Pbcm phases is very important, but it has fundamental limitations due to the nature of diffraction methods. Both, the powder XRD or electron diffraction do not allow us to unambiguously distinguish the Pca2₁ phase from the Pbca/Pbcm phases. It should be noted that the presence of ferroelectricity in thin $Hf_xZr_{1-x}O_2$ films can studied in detail using modern STEM methods, such as iDPC-STEM, dDPC and 4D-STEM (see e.g., Refs.[46, 47, 48]). These methods allow determining the exact positions of light atoms (in particular, oxygens), which makes it possible to analyze polarization processes and phase transitions. However, such studies for nanopowder samples are very complex, require a separate experimental design and modeling, which is beyond the scope of this work.

The average size (~5 - 10 nm) of the Hf$_x$Zr$_{1-x}$O$_2$ nanoparticles was determined by the high-resolution transmission electron microscopy (TEM) (see images in **Fig. 1(a)-(d)** as typical examples). Typical differential thermal analysis (DTA) of pyrolysis products is shown in **Fig. 1(e)**.

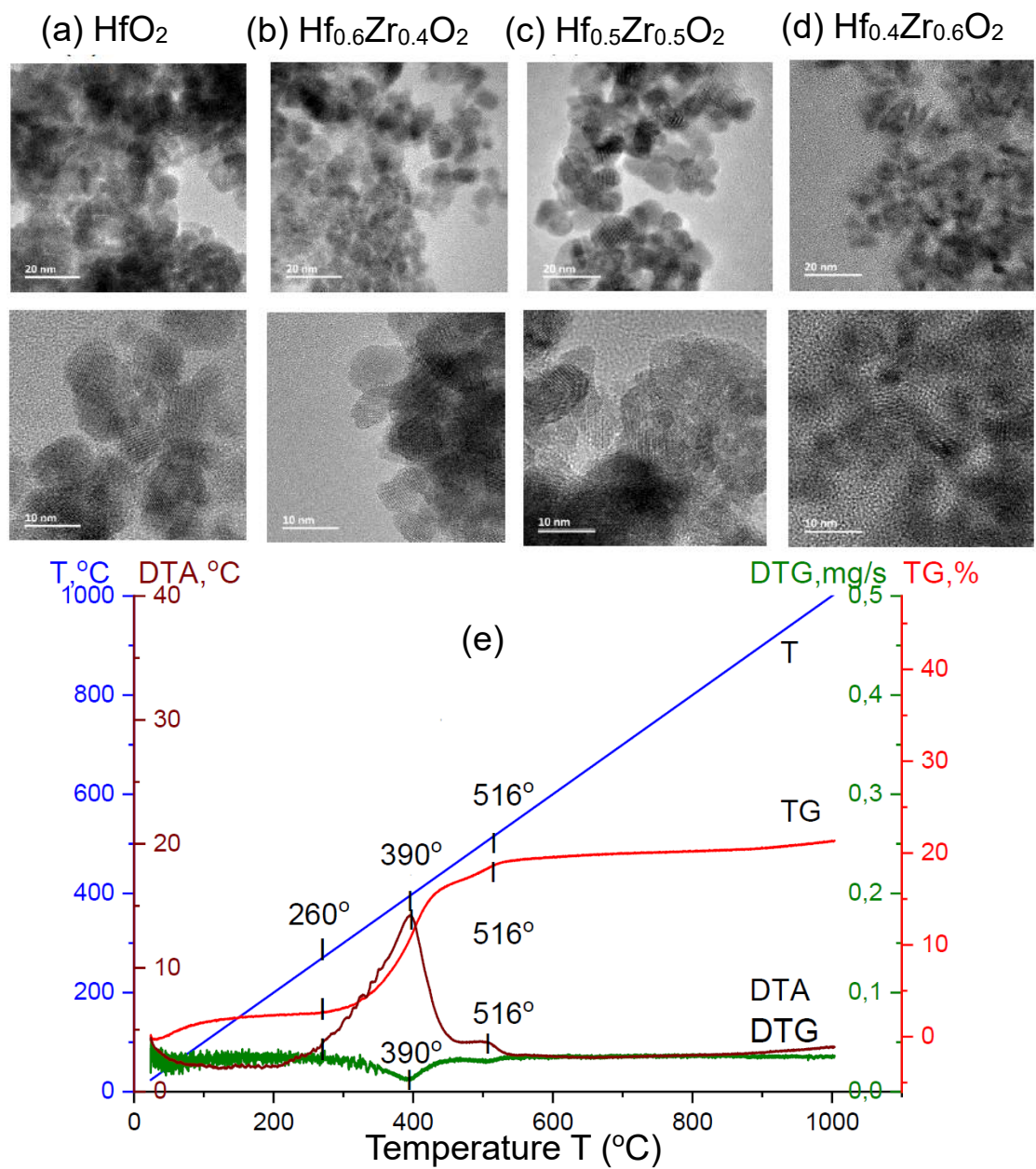

**FIGURE 1. (a-d)** Typical high resolution TEM images of the Hf$_x$Zr$_{1-x}$O$_2$ nanopowders. **(e)** DTA of pyrolysis products of a mixture of zirconium and hafnium nitrates (60:40) with fibrous cellulose at 600°C, 6 hours, in a CO+CO2 environment. Heating was performed in the air with a rate of 10 C/min.

For comparison with Hf$_x$Zr$_{1-x}$O$_2$ nanopowders, BaTiO$_3$ nanoparticles with the average size 24 nm and with the content of the ferroelectric phase 98 wt. % were prepared by the non-isothermal solid-state synthesis. Their preparation details are listed in **Appendix S1,** and their XRD characterization is given in Ref. [49].

## III. RESULTS OF ELECTRICAL MEASUREMENTS AND THEIR ANALYSIS

### A. Dielectric response and conductivity of the pressed $Hf_xZr_{1-x}O_2$ nanopowders

To study the transport of electric charge carriers in $Hf_xZr_{1-x}O_2$ nanopowders, the tableted powder samples were placed in a Teflon cylinder between two brass plungers, which serve to create uniaxial pressure and are used as electrical contacts. The distance between the plungers was determined by the thickness of the sample. The diameter of the samples is 4 mm, their thickness and the distance between the contacts are $(300 \pm 20)$ μm. Measurements of the capacitance and resistance of $Hf_xZr_{1-x}O_2$ nanopowders pressed at 5 MPa were carried out by the RLC-meter LCX200 ROHDE & SCHWARZ in the frequency range 4 Hz – 500 kHz.

The porosity of the studied samples was estimated as $(20 - 30)$ % in dependence on the applied pressure. We tried to account for the porosity under theoretical interpretation of the obtained results by using the effective media model. Within the model the pressed nanopowder is considered as a binary mixture of quasi-spherical core-shell $Hf_xZr_{1-x}O_2$ nanoparticles (with volume fraction μ) and air (with volume fraction 1-μ). The simplest way to account for the porosity is to regard μ as a fitting parameter that varies in a reasonable range for the fitting of the effective dielectric response, which is measured below and analyzed theoretically in Section IV.A.

The temperature dependences of the effective relative dielectric permittivity and resistivity at different frequencies for the pressed $Hf_xZr_{1-x}O_2$ nanopowders are shown **Fig. 2.** In the frequency range from 4 Hz to 1 kHz the effective permittivity has a pronounced maximum at 38°C for the $HfO_2$ sample, a pronounced maximum at 38°C followed by the quasi-plateau up to 88°C for the $Hf_{0.6}Zr_{0.4}O_2$ sample, a relatively sharp Λ-shaped maximum at 50°C for the $Hf_{0.5}Zr_{0.5}O_2$ sample, and a high maximum at 88°C with a left-side shoulder at 38°C for the $Hf_{0.4}Zr_{0.6}O_2$ sample. The maximum height decreases strongly and monotonically with an increase in frequency from 4 Hz to 500 kHz. At low frequencies (~4 Hz) the maximum height is about $1.5 \cdot 10^3$ for $HfO_2$ nanopowder, $2 \cdot 10^3$ for $Hf_{0.6}Zr_{0.4}O_2$ nanopowder, $1.1 \cdot 10^4$ for $Hf_{0.5}Zr_{0.5}O_2$ nanopowder and about $1.3 \cdot 10^5$ for $Hf_{0.4}Zr_{0.6}O_2$ nanopowder. The increase in the maximum height correlates with a decrease in Hf content from 1 to 0.4. At high frequencies (~500 kHz) the maxima height is much smaller; it changes from 7 (for $HfO_2$ nanopowder) to 20 (for $Hf_{0.4}Zr_{0.6}O_2$ nanopowder). The frequency dispersion of the dielectric permittivity maximum position is almost absent, meanwhile the shape and width of the maximum changes in a complex way with the frequency change. The position of the dielectric permittivity maximum coincides with the pronounced minimum of the electric resistivity for all studied samples.

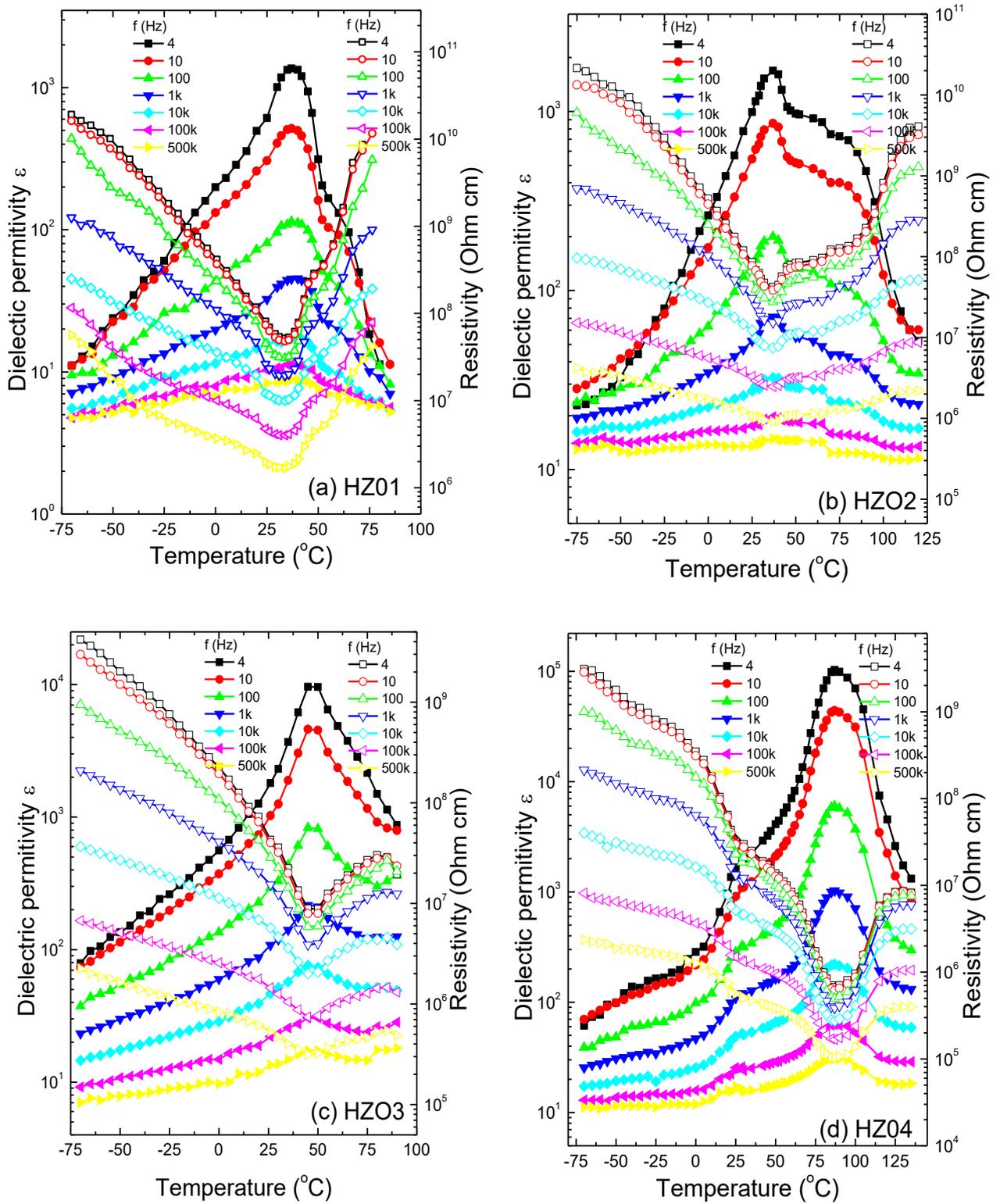

**FIGURE 2.** Temperature dependences of the relative dielectric permittivity (filled symbols) and resistivity (empty symbols) of the pressed HfO$_2$ **(a)**, Hf$_{0.6}$Zr$_{0.4}$O$_2$ **(b),** Hf$_{0.5}$Zr$_{0.5}$O$_2$ **(c)** and Hf$_{0.4}$Zr$_{0.6}$O$_2$ **(d)** nanopowders at different frequencies from 4 Hz to 500 kHz. The amplitude of the test sinusoidal voltage is 100 mV.

The temperature and frequency behavior of the Hf$_x$Zr$_{1-x}$O$_2$ nanopowders permittivity and resistivity is qualitatively similar to behavior of these dependences in the pressed BaTiO$_3$ nanopowder, shown in **Fig. 3.** It is seen that the permittivity of BaTiO$_3$ nanopowders has a pronounced maximum at 73°C, which corresponds to the ferroelectric-paraelectric phase transition and to related changes from the tetragonal to the cubic symmetry. The transition is shifted to lower temperatures in comparison with a bulk BaTiO$_3$ (~125°C) due to the size effects in stress-free small nanoparticles [50], while anisotropic strains in the core can lead to the opposite trend [51]. The maximum height decreases strongly and monotonically with an increase in frequency from $4 \cdot 10^5$ (at 4 Hz) to 30 (at 500 kHz). The frequency dispersion of the maximum position is almost absent, but its shape changes with frequency. The position of the dielectric permittivity maximum coincides with the pronounced minimum of the electric resistivity.

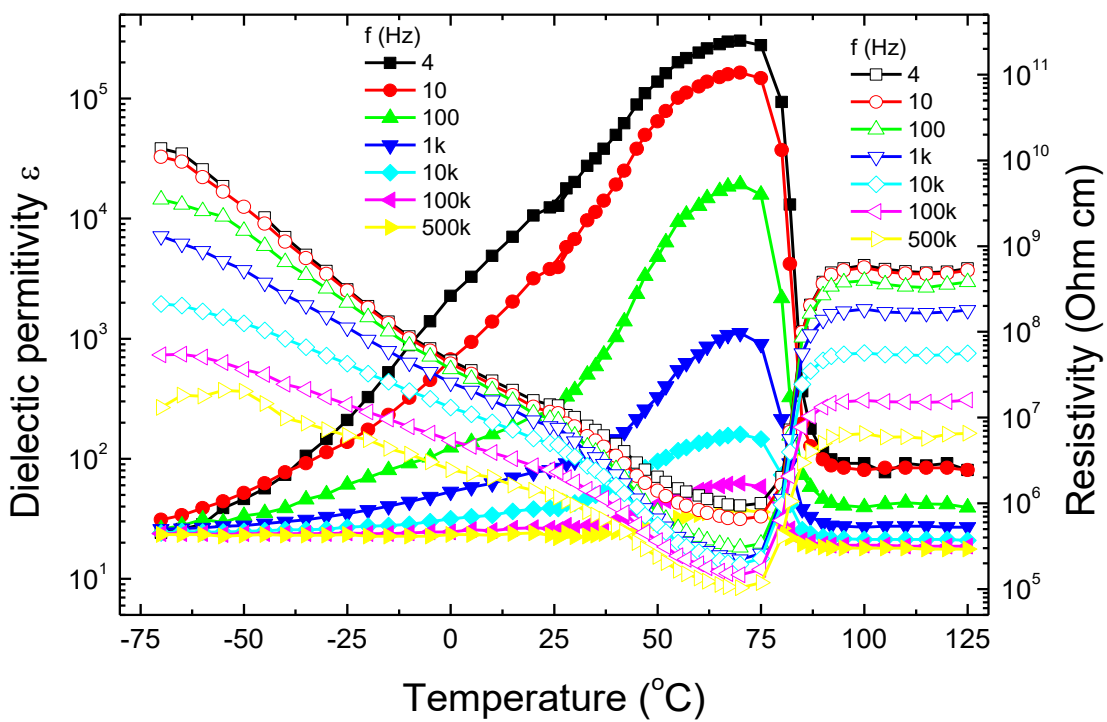

**FIGURE 3.** Temperature dependences of the relative dielectric permittivity (filled symbols) and resistivity (empty symbols) of the pressed BaTiO$_3$ nanopowder at different frequencies from 4 Hz to 500 kHz.

Note that non-polar bulk dielectrics HfO$_2$ and ZrO$_2$ have relative dielectric permittivity about 25 [52]. The dielectric permittivity of thin Hf$_{0.5}$Zr$_{0.5}$O$_2$ films [53] lies in the range 30-50; it may reach 80 for thin HfO$_2$ films with La, Bi or Nb substitutions [54], being almost temperature independent. At the same time, it is known that for typical perovskite ferroelectrics like BaTiO$_3$ and PbTiO$_3$ the relative dielectric permittivity depends strongly on temperature (reaching a sharp maximum ~$10^3$ – $10^4$ at Curie temperature). It also depends on the structure and morphology of the ferroelectric being different for single crystals, nanopowders and fine-grained ceramics. The experimentally measured dielectric permittivity of PbTiO$_3$ powders [55, 56, 57] varies from 100 to 1200 at room temperature (being size dependent). The dielectric permittivity of the fine-grained $BaTiO_3$ ceramics [58, 59] and $BaTiO_3$ nanoparticles [60] varies in the range 1000 – 7000 at room temperature.

The comparison of permittivity and resistivity behavior in $Hf_xZr_{1-x}O_2$ and $BaTiO_3$ nanopowders allows us to suspect a ferroelectric-like phase transition in the $Hf_xZr_{1-x}O_2$ nanopowders, which occur in the temperature range from 38°C (for $HfO_2$ nanopowder) to 88°C (for $Hf_{0.4}Zr_{0.6}O_2$ nanopowder). Notably that the temperature of the possible transition increases monotonically with increase in Zr content, meanwhile the shape and width of the dielectric permittivity maximum changes in a more complex way.

However, a colossal dielectric permittivity of ferroelectric nanocomposites and nanograined ceramics is often related to the dipolar relaxation enhanced by interfacial polarization dynamics, leading to the interfacial barrier-layer capacitance (IBLC) effect [61], as well as by the inhomogeneous layers between the electrodes and the sample, known as the surface barrier layer capacitance (SBLC) effect [62]. The origin of colossal dielectric permittivity observed in $BaTiO_3$ nanopowders may be related to the synergy of IBLC and SBLC effects [61, 62, 63], interfacial polarization at the interior of insulating grain boundaries, and/or polaron hopping in semiconducting cores with a large number of induced charge carriers (see e.g., Refs. [64, 65]).

As a rule, the colossal dielectric permittivity related with the IBLC and/or SBLC effects is accompanied by an increase in conduction losses, being a typical manifestation of the Maxwell-Wagner (MW) effect [66]. The colossal values of their dielectric permittivity are accompanied by an increase in the conduction losses. A strong frequency dispersion of the features of dielectric permittivity and losses, namely the strong frequency shift of the maxima, minima and/or plateau positions, is inherent to the IBLC, SBLC and polaronic effects. Since the frequency dispersion of the dielectric permittivity maximum and resistivity minimum is absent in the studied $Hf_xZr_{1-x}O_2$ nanopowders, as well as the temperature of the permittivity maximum coincides with the temperature of the resistivity minimum entire the frequency range (see **Fig. 2**), it allows us to assume a signification contribution of the expected ferroelectric-like phase transition to the dielectric response.

However, we realize that this argumentation is not fully conclusive. Impedance spectroscopy analysis could help decouple contributions of ferroelectric grains from their IBLC and/or SBLC effects originated from grain boundaries and electrodes. Typical Nyquist plots are analyzed in **Supplementary Materials** [41] (see **Appendix S3** therein). Indeed, we observed the deviations of Nyquist plots from the ideal semicircle, characteristic for the simple RC-circuit connected in parallel. In our case the semicircle is squashy-shaped and asymmetric. Its pronounced right "shoulder" can be related with a chaotic network of connected capacitors and resistors. Therefore, the sample volume may be simulated by a chaotic network of capacitors and resistors, which conduction obeys, in principle, the percolation theory. Such deviations can be described in the framework of effective medium approximation, which are considered in detail **Section IV.A**.

Not less important is that we do not observe the frequency dispersion of the relative dielectric permittivity maximum in the $Hf_xZr_{1-x}O_2$ nanopowders, which is inherent to relaxor ferroelectrics [67, 68] and dipolar glasses [69, 70]. However, the maxima shape is diffuse or plateau-like, especially for $Hf_{0.6}Zr_{0.4}O_2$ nanopowders, which may indicate on the relaxor-like behavior [71]. Following Rivera et al. [72], we can suppose a diffuse ferroelectric-like phase transition in $Hf_xZr_{1-x}O_2$ nanopowders, which classification is much wider in comparison with relaxor ferroelectrics.

However, without direct proofs, such as piezoelectric force microscopy (PFM) response or polarization-field (P-E) hysteresis loops, the claims about ferroelectricity of oxygen-deficient $Hf_xZr_{1-x}O_2$ nanoparticles remain speculative.

To the best of our knowledge, the local piezoresponce loops, observed by the PFM, are informative for individual nanoflakes with the lateral size much larger than the tip size (5 – 50 nm). The studied $Hf_xZr_{1-x}O_2$ nanoparticles, which can be ferroelectric-like according to our claims, have quasi-spherical shape with the actual size of 5 – 15 nm. Larger $Hf_xZr_{1-x}O_2$ particles become non-ferroelectric according to theory [36, 38] and experiment for films [1]. Therefore, the PFM of individual quasi-spherical $Hf_xZr_{1-x}O_2$ nanoparticles with sizes 5 – 15 nm is a very difficult task, which unlikely will give us informative results.

The P-E hysteresis loops are not informative to verify the ferroelectricity in pressed nanopowders, if the relative dielectric permittivity of nanoparticles is 10 or larger, as in the considered case. In this case the electric field is concentrated in the air gaps between the particles and does not penetrate inside them. In result one requires to apply giant voltages (higher than the air breakdown voltage) to the capacitor with the nanopowder to reverse the polarization of the individual particles.

Pyroelectric-type measurements of the particles' state can give us direct qualitative evidence of ferroelectricity in the studied $Hf_xZr_{1-x}O_2$ nanopowders. Imagine a nanopowder with individual ferroelectric nanoparticles. Due to their small size (5 – 15 nm) the nanoparticles are single-domain at temperatures below the Curie temperature $T_C$. Assuming that the distribution of particle sizes is narrow and thus the scattering of their Curie temperature is small, the surface of polarized nanoparticles accumulates screening charges from the ambient medium below $T_C$. These charges screen the bound charge of the individual particles, whose spontaneous polarization can have different orientation. The charge accumulation can be fast or slow, but the screening charge should disappear above $T_C$ due to the bound charge disappearance. Thus, the maximal values of charge-discharge currents of polarized nanoparticles and the local discharges between the nanoparticles can be expected near $T = T_C$, as well as the sharp decrease of the charge accumulation ability of the system should happen above $T_C$ [73, 74]. To verify the hypothesis, we measured time dependences of the charging current at fixed DC voltage across ta sample (2 V) at different temperatures from 20 to 100°C (see

**Appendix S3** in Supplementary Materials [41]). The total charge $\Delta Q$ accumulated in the $Hf_{0.5}Zr_{0.5}O_2$ sample as a function of temperature is shown in **Fig. S1(h)** therein. It appears that the electric charge accumulated by the pressed nanopowders decreases strongly above 70°C. We related the observed decrease to the disappearance of the pyroelectric-type charge in the system above 70°C. For the $Hf_{0.5}Zr_{0.5}O_2$ sample, the accumulated charge is maximal in the temperature range 55 – 65°C, and the current is maximal near 50°C. Important, that the total discharge time (after the voltage removal) is more than 10 – 20 minutes, which may indicate the pronounced electret-like properties of the samples.

A quantitative description of the electric conduction mechanisms and dielectric response in the $Hf_xZr_{1-x}O_2$ nanopowders is given below.

### B. Electric conduction mechanisms in the alternate current regime

The frequency dependence of the $Hf_xZr_{1-x}O_2$ sample resistivity at temperatures below the phase transition (shown in **Fig. 4** for $Hf_{0.5}Zr_{0.5}O_2$), demonstrates that the samples manifest a gradual transition to dominating hopping conduction with a decrease in the temperature. It is seen that the frequency dependence at the lowest temperature is well fitted by the power law $\sigma \sim \omega^s$, where S<1, in the whole frequency range, which is characteristic for the hopping conduction at moderate frequencies [75, 76].

At the same time, the frequency dependence of resistivity tends to the lowering of its slope and becomes flatter at higher temperatures and lower frequencies. It evidences the gradual change of the hopping conduction to another mechanism caused by the change of crystallographic structure.

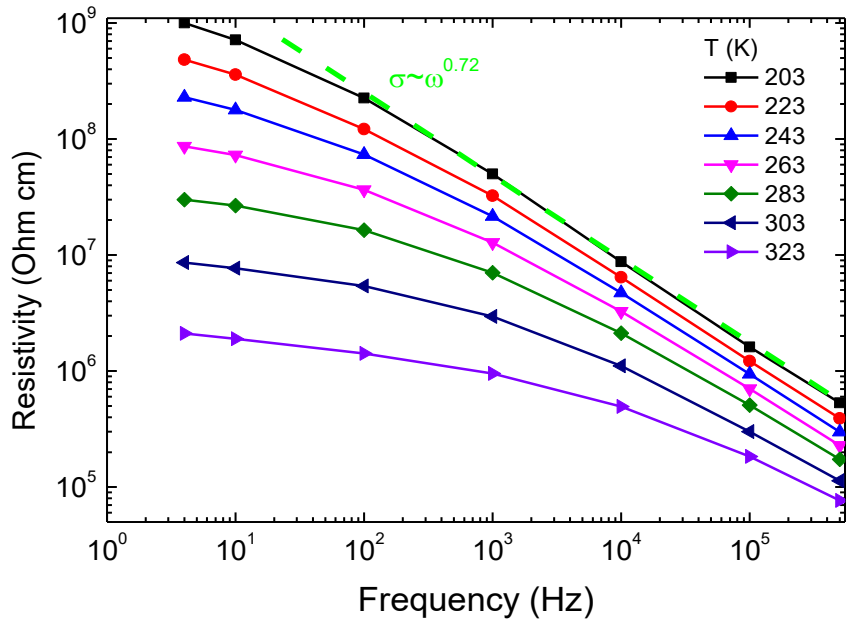

**FIGURE 4.** Frequency dependences of resistivity of the $Hf_{0.5}Zr_{0.5}O_2$ sample at different temperatures in the range below the phase transition temperature.

Shown in **Fig. 5(a)** are temperature dependences of resistivity of the $Hf_{0.5}Zr_{0.5}O_2$ sample at several highest frequencies in the Arrhenius coordinates. It is seen that below 290 K the dependences go out onto the activation type behavior with the activation energy of about 67 meV. Frequently, the low temperature conduction in the alternate current regime in the disordered materials, such as amorphous and doped semiconductors, granulose materials, etc., is explained based on the hopping conduction model [75, 77]. The hopping conduction in the mixed oxides is considered also within the frames of the small polaron model (see e.g. Refs. [78, 79]), which may give much larger activation energy as compared to the classical hopping model. Thus, one may presume that the high-frequency hopping conduction runs in the regime of small polaron hopping via the nearest neighbors which is characterized by the constant activation energy.

At lower frequencies the temperature dependences of resistivity do not reach a purely activated behavior, characteristic to any kind of hopping conduction (**Fig. 5(b)**). They do not fit by straight lines even in coordinates characteristic for the Mott variable range hopping conduction, $\rho \sim \exp\left[\left(\frac{T_0}{T}\right)^{1/4}\right]$ [80]. Presumably, such behavior may be related to more involvement in conduction of ionic interface conduction aside from the polarization mechanism.

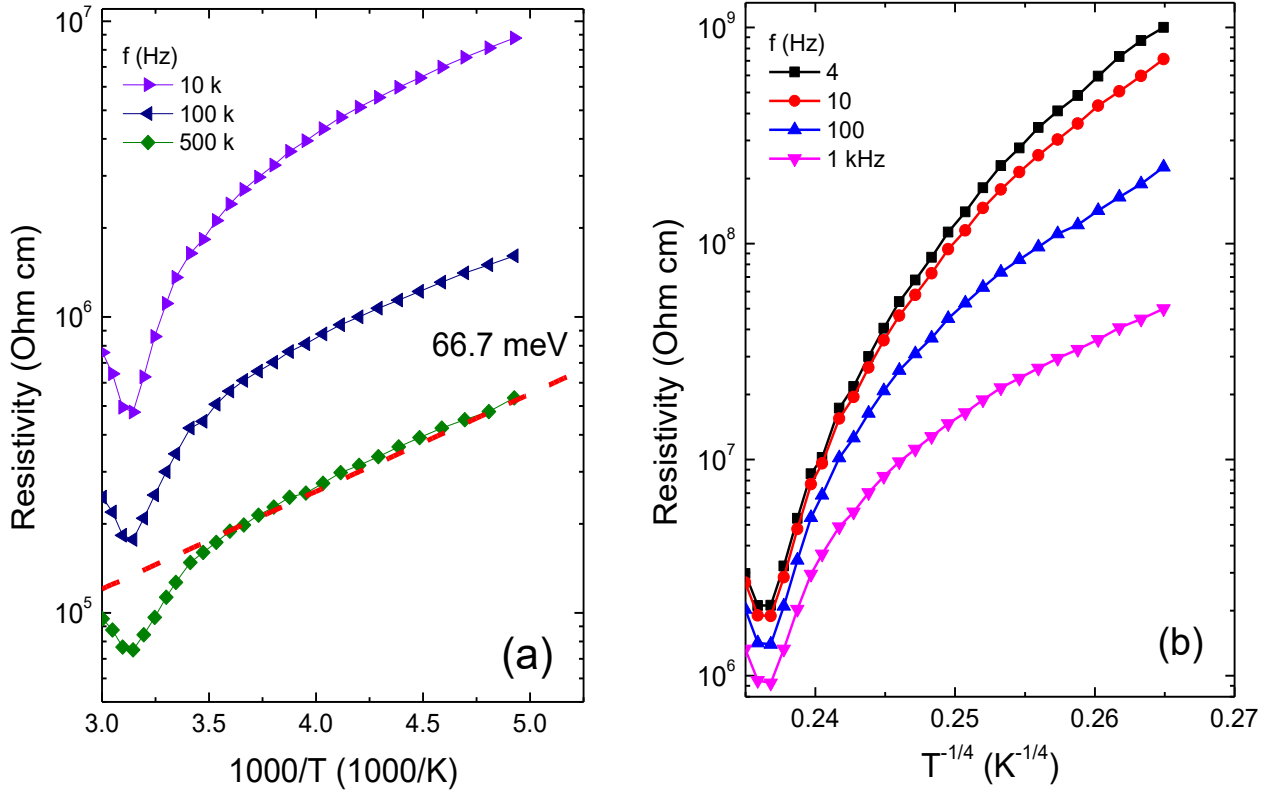

**FIGURE 5.** (a) Temperature dependences of the Hf$_{0.5}$Zr$_{0.5}$O$_2$ sample resistivity at three highest frequencies in the Arrhenius coordinates. (b) Temperature dependences of resistivity of the Hf$_{0.5}$Zr$_{0.5}$O$_2$ sample at several lowest frequencies in the coordinates of the Mott law: $\rho \sim \exp(T_0/T)^{1/4}$.

Among the models of electric conduction of ferroelectric oxide compounds not related to the hopping conduction is the model proposed by Heywang [40], which is based on the concept of creation of the interfacial potential Schottky barriers on the grain surfaces. The barrier height $\varphi_S$ is determined as

$$\varphi_S = \frac{e^2 n_D}{2\varepsilon_0 \varepsilon} d^2, \tag{1}$$

where $e$ is an elementary charge, $\varepsilon_0$ is a universal dielectric constant, $\varepsilon$ is a relative dielectric permittivity, $d$ is the width of the barrier layer, $n_D$ is the electron concentration providing the conduction due to presence of donor centers in the bulk. According to this model, one supposes that the resistance in the bulk is determined by electrons in the conduction band and obeys the law

$$R \sim \exp\left(\frac{\varphi_S}{k_B T}\right) = \exp\left(\frac{e^2 n_D d^2}{2\varepsilon_0 \varepsilon k_B T}\right) \cong \exp\left(\frac{A}{\varepsilon T}\right). \tag{2}$$

Accounting for the strong dependence of effective dielectric permittivity vs. temperature one may suppose this dependence to be fitted by the straight line in coordinates $\ln(R)$ and $\frac{1}{\varepsilon T}$.

However, as is seen in **Fig. 6**, this is not the case. The curve with blue symbols in Arrhenius coordinates, $\ln(R)$ and $\frac{1}{\varepsilon T}$, strongly decreases its slope. It may be understood supposing non-uniformity in size of nanograins constituting the powder, variation of concentration of centers determining the charge carrier concentration participating in the conductivity along with spread in polarizability among nanograins determining variation the surface barriers parameters. These changes apparently are over the area of a tabletted shape powder sample. It is interesting to note, that, as shown by the curve with black symbols in **Fig. 6**, the obtained dependence is quite well fitted by a straight line in coordinates $\ln(R)$ vs. $\left(\frac{1}{\varepsilon T}\right)^{1/4}$, which corresponds to proportionality $R \sim \exp\left[\left(\frac{A}{\varepsilon T}\right)^{1/4}\right]$ similar to the case of the variable range hopping (VRH) conduction by the Mott law [80]. This may prove the assumption of non-uniformity of the powder sample over its area.

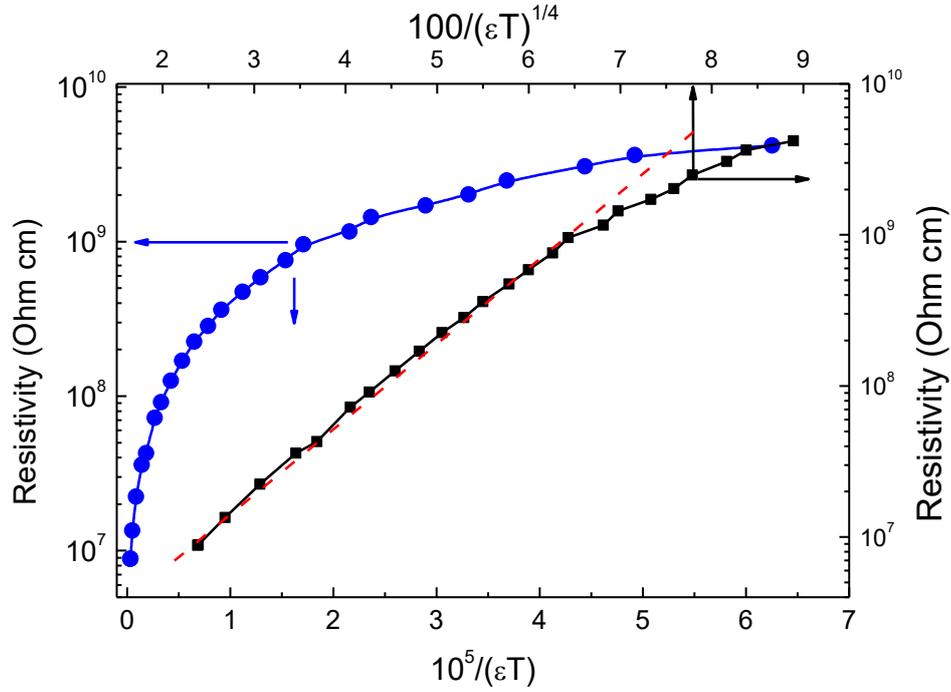

**FIGURE 6**. The temperature dependence of the Hf$_{0.5}$Zr$_{0.5}$O$_2$ sample resistivity at low frequency in coordinates modified according to the Heywang model [40]. The curve with blue symbols is plotted in the Arrhenius law coordinates $\ln(R)$ vs. $\frac{1}{\varepsilon T}$. The curve with black symbols is plotted in the Mott law coordinates $\ln(R)$ vs. $\left(\frac{1}{\varepsilon T}\right)^{1/4}$. The test signal frequency is 4 Hz.

One should also note the impact of the frequency increase on this model. Shown in **Fig. S6** [41] is the temperature dependence of resistivity for the Hf$_{0.5}$Zr$_{0.5}$O$_2$ sample in Arrhenius coordinates $R \sim \exp\left(\frac{A}{\varepsilon T}\right)$ at the test signal frequency of 500 kHz. Here the curve is closer to the Arrhenius behavior. At higher frequency the areas of the sample with the lower barrier energy and consequently lower resistivity dominate in the total resistivity.

Since we observed that the temperature of the pronounced minimum of resistivity coincides with the temperature of the maximal permittivity for both low and high frequencies (see **Fig. 7**), it allows to suspect the validity of the Heywang model [40] for the Hf$_x$Zr$_{1-x}$O$_2$ nanopowders. It seems very important that the temperature dependencies of the dielectric permittivity (upper curves in **Fig. 7**) and resistivity (bottom curves in **Fig. 7**) are almost mirror-like reflected in respect to each other, which means that all their features, such as position and shape of maxima, plateau, minima and inflexions, as well as straight parts, almost coincide after the mirror reflection in respect to the temperature axis. In other words, this means that $\ln(R) \sim \left(\frac{1}{\varepsilon T}\right)^\lambda$. A quantitative description of the observed temperature and frequency behavior of the permittivity and resistivity is given in the next section.

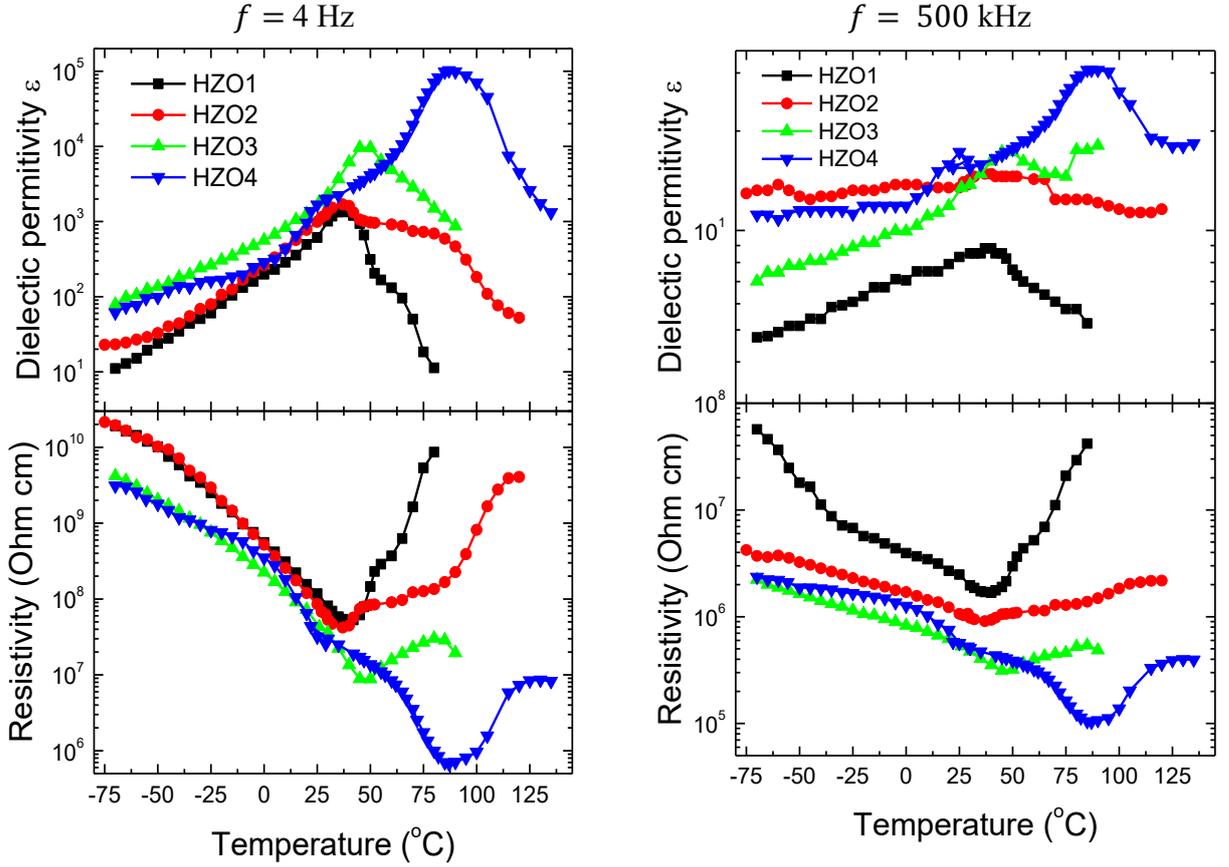

**FIGURE 7.** Temperature dependences of the dielectric permittivity (filled symbols) and resistivity (empty symbols) of the samples HZO1-4 at frequencies of 4 Hz (left) and 500 kHz (right).

Thus, one can conclude that the temperature dependence of resistivity at temperatures both below and above the Curie temperature may be explained in the framework of Heywang model [40] accounting for creations of interfacial barriers on the nanoparticle surface and impact of efficient dielectric permittivity on their height. The temperature dependence of resistivity in this case is well described by the Mott law for the VRH conduction model if we replace the temperature $T$ by the efficient temperature $T_{eff} = \varepsilon T$. In the case of adopting the hopping conduction model to explain our results, it may mean that the dielectric permittivity also contributes significantly to the modulation of the density of states (DOS) at the Fermi level. The dielectric permittivity decreases strongly and varies slowly at the lowest temperature. In this case the fingerprints of the hopping conduction appear also, which consist in the characteristic frequency dependence of $\omega^S$. Here one already does not need an efficient temperature to consider. Meanwhile this model still demands further development and additional experiments.

# IV. THEORETICAL MODELING OF THE DIELECTRIC RESPONSE

## A. Effective Medium Approximation

There are many effective medium approximations (shortly "EMA") [81], among which the most known are the Landau approximation of linear mixture [82], Maxwell-Garnett [83] and Bruggeman [84] approximations for spherical inclusions, and Lichtenecker-Rother approximation of the logarithmic mixture [85]. Most of these approximations are applicable for quasi-spherical randomly distributed dielectric (or semiconducting) particles in the insulating environment. These EMA models lead to analytical expressions for the effective permittivity $\varepsilon^*_{eff}$ [86]. However, their applicability ranges are very sensitive to the cross-interaction effects of the polarized ferroelectric particles, and therefore most of them are invalid for ferroelectric mixtures and/or colloids, where the content of the particles is more than (20–30) vol.% [87].

EMA models for arbitrary content of the nanoparticles were developed by Petzelt et al. [88] and Rychetský et. al. [89]. As a rule, EMA considers the algebraic equation for the effective permittivity of the binary mixture:

$$(1-\mu)\frac{\varepsilon^*_{eff}-\varepsilon^*_b}{(1-n_a)\varepsilon^*_{eff}+n_a\varepsilon^*_b} + \mu\frac{\varepsilon^*_{eff}-\varepsilon^*_a}{(1-n_a)\varepsilon^*_{eff}+n_a\varepsilon^*_a} = 0. \qquad (3)$$

Here $\varepsilon^*_a$, $\varepsilon^*_b$ are complex functions of the relative permittivity of the components "$a$" and "$b$" respectively, $\mu$ and $1-\mu$ are relative volume fractions of the components "$a$" and "$b$" respectively, $n_a$ is the depolarization field factor for the inclusions of the type "$a$". In the case $\mu=0$ or $\mu=1$, the solution of Eq.(3) is $\varepsilon^*_{eff} = \varepsilon^*_b$ or $\varepsilon^*_{eff} = \varepsilon^*_a$, respectively.

For $n_a = 1$ (i.e., for the system consisting of the layers, perpendicular to the external field) one reduces Eq.(3) to the expression for the Maxwell layered dielectric model, $\varepsilon^*_{eff} = \left(\frac{1-\mu}{\varepsilon^*_b} + \frac{\mu}{\varepsilon^*_a}\right)^{-1}$. For $n_a = 1/3$ (i.e., for the system consisting of the spherical nanoparticles) one rewrites (3) as $\varepsilon^*_{eff} = \varepsilon^*_b\left[1 + \frac{3\mu(\varepsilon^*_a-\varepsilon^*_b)}{\varepsilon^*_a+2\varepsilon^*_b-\mu(\varepsilon^*_a-\varepsilon^*_b)}\right]$, which is the Wagner expression.

For $n_a = 0$ (i.e., for the system of the columns, parallel to the external field) Eq.(3) yields $\varepsilon^*_{eff} = (1-\mu)\varepsilon^*_b + \mu\varepsilon^*_a$, which is equivalent to the system with parallel connected capacitors with the complex permittivity $\varepsilon^*_b$ and $\varepsilon^*_a$. The columnar model allows the separate fitting of the real ($\varepsilon_{eff}$) and imaginary ($\sigma_{eff}$) parts of $\varepsilon^*_{eff}$, which is the simplest case considered hereinafter among others.

Following EMA, we assume that the cores of Hf$_x$Zr$_{1-x}$O$_2$ nanoparticles (abbreviated as "C") and their non-ferroelectric shells (abbreviated as "S") have a complex dielectric permittivity:

$$\varepsilon^*_C(T,\omega) = \varepsilon_C(T,\omega) - i\frac{\sigma_C(T,\omega)}{\varepsilon_0\omega}, \qquad \varepsilon^*_S(T,\omega) = \varepsilon_S(T,\omega) - i\frac{\sigma_S(T,\omega)}{\varepsilon_0\omega}. \qquad (4a)$$

The effective dielectric permittivity is

$$\varepsilon_{eff}^*(T,\omega) = \varepsilon_{eff}(T,\omega) - i\frac{\sigma_{eff}(T,\omega)}{\varepsilon_0 \omega}. \quad (4b)$$

Following Heywang model [40] and considering experimental results shown in **Fig. 7**, the expected Arrhenius-type and/or the Mott-type temperature dependences (as well as a general stretched-exponential law) for the effective conductivity of the nanopowders could be modified by introduction of the effective dielectric permittivity. Thus, we use the following fitting function for effective conductivity:

$$\sigma_{eff}(T,\omega) = \sigma_A^0(\omega)\exp\left[-\left(\frac{E_A}{k_B T \varepsilon_{eff}(T)}\right)^\lambda\right]. \quad (5)$$

Here $E_A$ is an activation energy of the space charges in the cores/shells (A = C or S). The positive fitting parameter $\lambda$ varies in the range $0 < \lambda \leq 1$ for a general stretched-exponential law, which includes the Mott law with $\lambda = 1/4$ and the Arrhenius law with $\lambda = 1$. Varying $\lambda$ one can simulate a possible crossover from the Arrhenius-type dependence to the Mott-type dependence of the effective conductivity.

Hereinafter we assume that $\varepsilon_{eff}(T)$ is a real positive value corresponding to the static (i.e., low-frequency) effective dielectric permittivity of the mixture, which should be determined in a self-consistent way or taken from the experiment. It is important that we use the effective value $\varepsilon_{eff}(T)$ in Eqs.(5), which can be determined from experiment, instead of unknown local values of the dielectric permittivity. The assumption explains the minimum of resistivity near the temperature maximum of $\varepsilon_{eff}(T)$.

Fitting of the effective resistivity based on Eq.(5) is shown in **Fig. 8**. It appeared that the Heywang model [40] can describe the experimental results quantitatively for the power $\lambda_C \approx \lambda_S \approx \lambda = 1/4$ for both low and high frequencies (see dashed curves in **Fig. 8** and **Table II**), which corresponds the Mott-type VRH conduction in the $Hf_xZr_{1-x}O_2$ nanopowders. At the same time the best fitting corresponds to $\lambda = 1/9$ (see solid curves in **Fig. 8** and **Table II**), but the difference between the fitting curves for $\lambda = 1/9$ and $\lambda = 1/4$ is significant at temperatures well below the temperature maximum of $\varepsilon_{eff}(T)$. At low frequences the core contribution dominates, and the shell contribution increases with an increase in frequency (or vice versa).

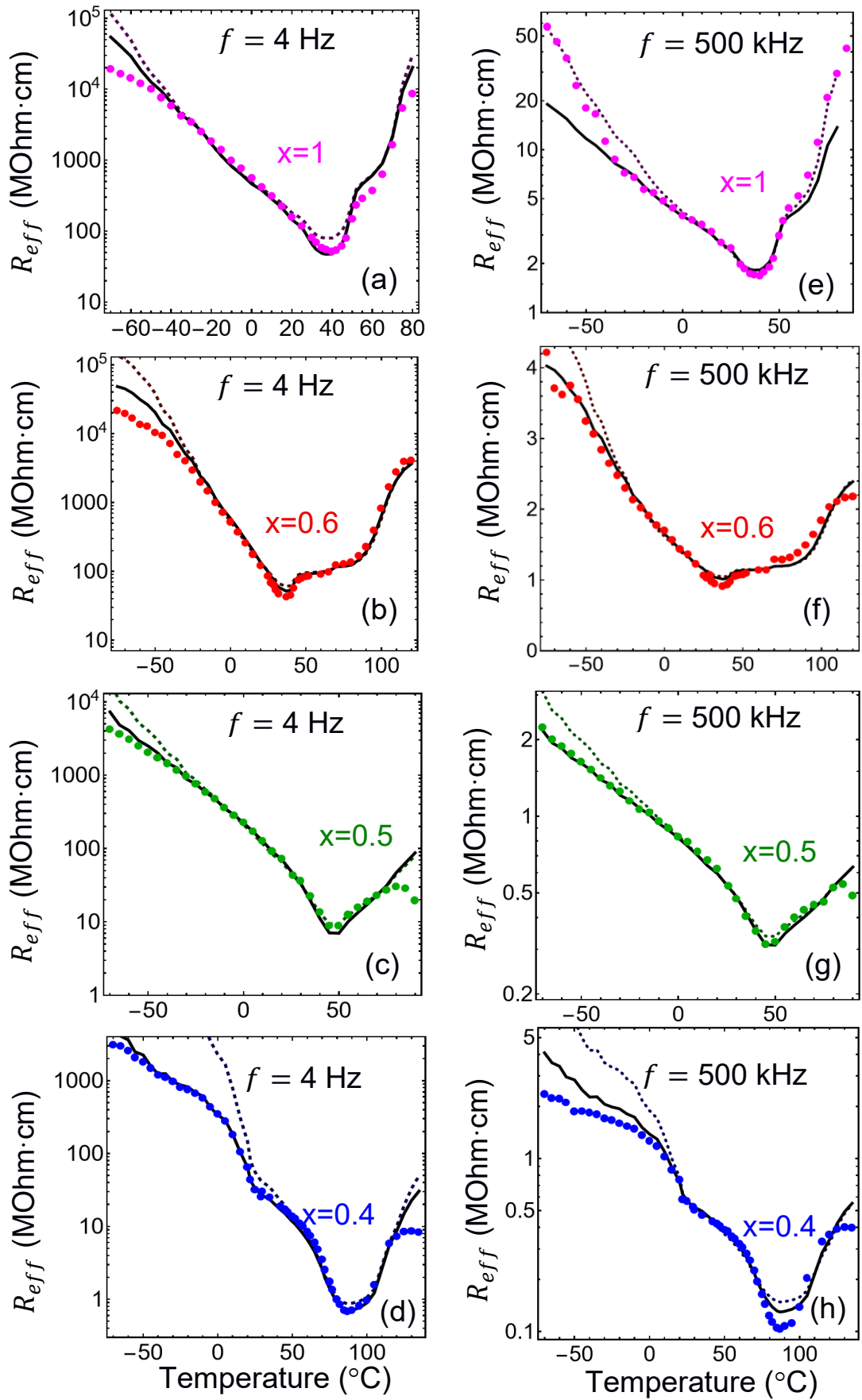

**FIGURE 8**. Temperature dependences of the effective resistivity of the $Hf_xZr_{1-x}O_2$ nanopowders with different Hf content x=1 **(a, e)**, 0.6 **(b, f)**, 0.5 **(c, g)** and 0.4 **(d, h)**. Symbols are experimentally measured values with the test signal frequency 4 Hz **(a, b, c, d)** and 500 kHz **(e, f, g, h)**. Curves are the fitting by expression $R \cong R_0 \exp\left[(A/\varepsilon_{eff}T)^\lambda\right]$, for $\lambda = 1/9$ (solid curves) and $\lambda = 1/4$ (dashed curves). The values of $\varepsilon_{eff}$ are taken from experimental data at 4 Hz. The fitting parameters $R_0$ and $A$ are listed in **Table II.**

**Table II**. The values of the fitting parameters $R_0$ and $A$ used to describe experimentally measured resistivity

| Sample | x | Frequency $f$ =4 Hz | | Frequency $f$ =500 kHz | |
|---|---|---|---|---|---|
| | | $\lambda = 1/4$ | $\lambda = 1/9$ | $\lambda = 1/4$ | $\lambda = 1/9$ |
| HZO1 (HfO$_2$) | 1 | $A = 2.230 \times 10^7$ $R_0 = 5.355 \times 10^6$ | $A = 1.529 \times 10^{14}$ $R_0 = 6.220 \times 10^3$ | $A = 1.123 \times 10^6$ $R_0 = 4.920 \times 10^5$ | $A = 7.662 \times 10^9$ $R_0 = 9.326 \times 10^4$ |
| HZO2 (Hf$_{0.6}$Zr$_{0.4}$O$_2$) | 0.6 | $A = 6.845 \times 10^7$ $R_0 = 2050 \times 10^6$ | $A = 4.517 \times 10^{14}$ $R_0 = 2.756 \times 10^3$ | $A = 1.086 \times 10^5$ $R_0 = 5.326 \times 10^5$ | $A = 2.491 \times 10^8$ $R_0 = 1.394 \times 10^5$ |
| HZO3 (Hf$_{0.5}$Zr$_{0.5}$O$_2$) | 0.5 | $A = 1.854 \times 10^8$ $R_0 = 5.810 \times 10^5$ | $A = 9.196 \times 10^{14}$ $R_0 = 1.121 \times 10^3$ | $A = 1.354 \times 10^6$ $R_0 = 1.496 \times 10^5$ | $A = 9.692 \times 10^9$ $R_0 = 2.701 \times 10^4$ |
| HZO4 (Hf$_{0.4}$Zr$_{0.6}$O$_2$) | 0.4 | $A = 7.870 \times 10^8$ $R_0 = 1.014 \times 10^5$ | $A = 5.808 \times 10^{14}$, $R_0 = 1.267 \times 10^3$ | $A = 0.859 \times 10^7$ $R_0 = 7.387 \times 10^4$ | $A = 1.007 \times 10^{11}$ $R_0 = 1.167 \times 10^4$ |

\* Parameter $A$ in Kelvin, $R_0$ in Ohm·cm

Hereinafter, we assume that the shell dielectric permittivity $\varepsilon_S$ monotonically and relatively weakly depends on temperature, but can be frequency-dependent:

$$\varepsilon_S(T,\omega) \approx \varepsilon_b^0(\omega). \qquad (6a)$$

Next, we assume that the dielectric permittivity of the cores consists of the dielectric-like (or paraelectric-like) contribution, originating from the non-ferroelectric modes, and the ferroelectric-like contribution, which can obey the Curie-Weiss type law. For the displacive-type ferroelectric mode the dielectric susceptibility $\chi$ obeys the following phenomenological equation, $\chi = \frac{1}{\alpha(T)+3\beta P_3^2+5\gamma P_3^4+7\delta P_3^6}$, where $\alpha$ $\beta$, $\gamma$, and $\delta$ are the Landau expansion coefficients of the free energy over the even powers of the order parameter – polarization $P_3$ [38, 51]. According to the LGD approach, the following fitting function for the permittivity $\varepsilon_C(T,\omega)$ of the core with a diffuse phase transition can be used:

$$\varepsilon_C(T,\omega) = \varepsilon_b^0(T,\omega) + \frac{C_W}{\sqrt{\left(T-T_C+3\vartheta|T-T_C|L^2(T)\right)^2+\Delta_d^2(\omega)}}. \qquad (6b)$$

Here $\varepsilon_b^0(T,\omega)$ is the non-ferroelectric "background" permittivity, which can be constant or linear function of temperature. The positive constant $C_W$ is the analog of the Curie-Weiss constant, $T_C$ is the Curie temperature of the ferroelectric-like transition, $\Delta_d^2$ is the permittivity dispersion and $\xi$ is the power of the permittivity decay. The fitting parameter $\vartheta$ can vary in the range $\frac{1}{3} < \vartheta \leq 3$. The dimensionless polarization parameter $L(T) = \frac{P_3(T)}{P_S(0)}$ is approximated as $L(T) \approx \frac{1}{1+\exp\left(\frac{T-T_C}{\Delta_c(\omega)}\right)}$, where the dispersion $\Delta_c(\omega)$ models the diffuseness of the phase transition. The condition $\Delta_c(\omega) \ll T_C$ corresponds to a relatively sharp transition at $T = T_C$. Note that $\vartheta = 1$, and $\Delta_d^2 \to 0$ for the classical Curie-Weiss law in displacive ferroelectrics with the second order phase transition.

For the case of the order-disorder type ferroelectric core, where the order parameter satisfies the equation, $L(T) = \tanh\left[\frac{T_C}{T} \cdot L(T)\right]$, the following fitting function for the core permittivity $\varepsilon_C(T, \omega)$ can be used:

$$\varepsilon_C(T, \omega) = \varepsilon_b^0(\omega) + \frac{C_W}{T \cosh^2\left[\frac{T_C}{T} \cdot L(T)\right] - T_C}. \tag{6c}$$

The temperature dependence of the dimensionless order parameter $L(T)$ can be approximated as in the previous case, $L(T) \approx \frac{1}{1+\exp\left(\frac{T-T_C}{\Delta_c(\omega)}\right)}$, where the dispersion $\Delta_c(\omega) \ll T_C$ for the sharp transition. Here the background constant $\varepsilon_b^0(\omega)$, the Curie-Weiss constant $C_W$ and the dispersion $\Delta(\omega)$ are fitting parameters.

The fitting of the low-frequency and high-frequency dielectric permittivity using hypotheses (6b) and (6c) is shown in **Figs. S7 – S8** and **S9** [41], respectively. Using the Bayesian optimization with these hypotheses (see e.g., details in Refs. [90, 91]), we can estimate the probabilities of the displacement-type and order-disorder type ferroelectric-like transitions in the $Hf_xZr_{1-x}O_2$ nanopowders.

Temperature dependences of the effective dielectric permittivity $\varepsilon_{eff}(T, \omega)$ of the $Hf_xZr_{1-x}O_2$ nanopowders with different Hf content are shown in **Fig. 9**. Solid curves are the fitting using Eqs.(6b) and Bayesian optimization. Fitting parameters are listed in **Table III.** It appeared that the best fitting corresponds to significant deviation from the classical Curie-Weiss law (with $\vartheta = 1$ and $\Delta_d \to 0$), at that all fitting parameters, except $\varepsilon_b^0$, depend significantly on the frequency $f$. The most significant are the frequency dependences of the Curie-Weiss constant $C_W$, dispersion $\Delta_d$ and parameter $\vartheta$ (see **Table III**). The parameters $C_W$ and $\Delta_d$ mostly increase strongly with an increase in frequency. The parameter $\vartheta$ decreases (from 30 % to 3 times) with an increase in frequency from 4 Hz to 500 kHz. The frequency dispersion of $T_C$ is relatively weak (from 0.1 K to 8 K) and may be ascribed to the fitting accuracy for the scattered data.

At the same time the main maximum of the dielectric permittivity is very large in comparison with other features for all compositions "x" (see **Figs. 9**), and its width is much larger in comparison with the ordered crystalline ferroelectrics. These trends agree with our assumption that the studied $Hf_xZr_{1-x}O_2$ nanopowders undergo a diffuse ferroelectric-like phase transition in the temperature range 38 – 98 °C.

However, we are aware that the colossal dielectric permittivity, observed at low frequencies in the $Hf_xZr_{1-x}O_2$ nanopowders, can originate from the IBLC and/or SBLC effects [61, 62] accompanied by an increase in conduction losses [63]. Equation (3), as the EMA model equation, describes the IBLC and/or SBLC effects, and, at the same time, includes the limiting cases of the Maxwell layered dielectric model ($n_a = 1$), the Wagner model of spherical nanoparticles ($n_a = 1/3$), and the columnar model ($n_a = 0$). According to the fitting results of the measured dielectric response and conductivity, based on the Bayesian optimization with the working hypotheses (5)-(6), the columnar model is the most probable. This result allows the separate fitting of the real and imaginary parts of $\varepsilon_{eff}^*$ from Eqs.(5)-(6) and predicts that the ferroelectric-like phase transition may take place in the semiconducting cores of $Hf_xZr_{1-x}O_2$ nanoparticles.

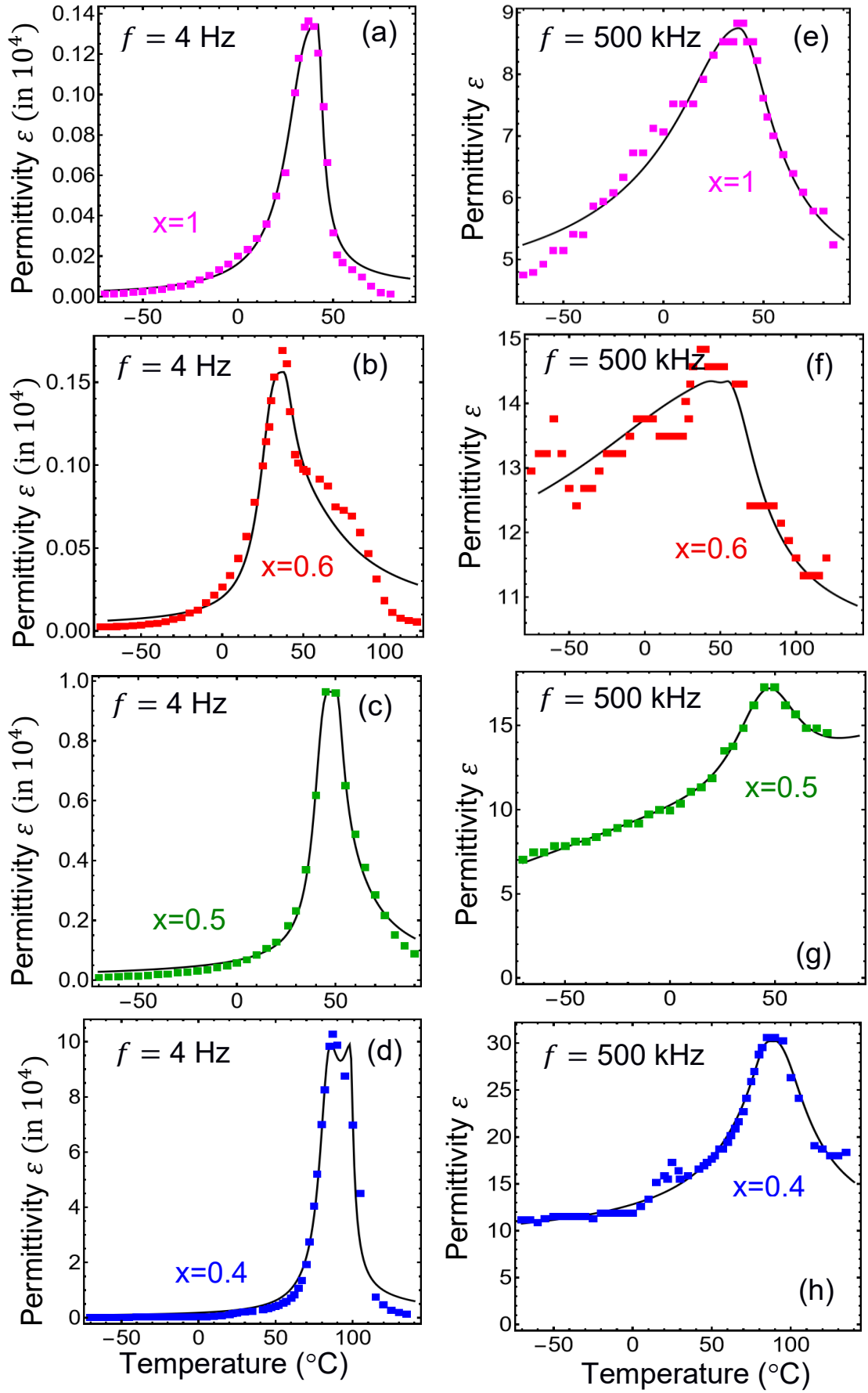

**FIGURE 9.** Temperature dependences of the effective dielectric permittivity of the $Hf_xZr_{1-x}O_2$ nanopowders with different Hf content x=1 **(a, e)**, 0.6 **(b, f)**, 0.5 **(c, g)** and 0.4 **(d, h)**. Symbols are experimentally measured values with the test signal frequency 4 Hz **(a, b, c)** and 500 kHz **(d, e)**. Black curves are calculated using the expressions (6a), (6b) and Bayesian optimization. The fitting parameters are listed in **Table III.**

**TABLE III.** The values of the fitting parameters in Eqs.(6a) and (6b) used to describe the effective dielectric permittivity

| Sample | x | Frequency $f = 4$ Hz | Frequency $f = 500$ kHz |
|---|---|---|---|
| HZO1 (HfO$_2$) | 1 | $\varepsilon_b^0 = 4$, $\Delta_d = 4.5$ K<br>$C_W = 6080.6$ K<br>$\Delta_c = 67.3$ K, $\nu = 1.50$<br>$T_C = 42.0$°C | $\varepsilon_b^0 = 4$, $\Delta_d = 15.2$ K<br>$C_W = 72.1$ K<br>$\Delta_c = 0.3$ K, $\nu = 0.50$<br>$T_C = 37.7$°C |
| HZO2 (Hf$_{0.6}$Zr$_{0.4}$O$_2$) | 0.6 | $\varepsilon_b^0 = 5$, $\Delta_d = 14.8$ K<br>$C_W = 23021.4$ K<br>$\Delta_c = 15.2$ K, $\nu = 1.57$<br>$T_C = 37.9$°C | $\varepsilon_b^0 = 5$, $\Delta_d = 61.3$ K<br>$C_W = 558.3$ K<br>$\Delta_c = 2.8$ K, $\nu = 0.45$<br>$T_C = 44.3$°C |
| HZO3 (Hf$_{0.5}$Zr$_{0.5}$O$_2$) | 0.5 | $\varepsilon_b^0 = 5$, $\Delta_d = 5.77$ K<br>$C_W = 55626.5$ K<br>$\Delta_c = 8.5$ K, $\nu = 0.89$<br>$T_C = 50.6$°C | $\varepsilon_b^0 = 5$, $\Delta_d = 30.03$ K<br>$C_W = 350.7$ K<br>$\Delta_c = 3.9$ K, $\nu = 0.74$<br>$T_C = 50.5$°C |
| HZO4 (Hf$_{0.4}$Zr$_{0.6}$O$_2$) | 0.4 | $\varepsilon_b^0 = 8$, $\Delta_d = 2.8$ K<br>$C_W = 277270.2$ K<br>$\Delta_c = 29.7$ K, $\nu = 0.91$<br>$T_C = 98.4$°C | $\varepsilon_b^0 = 8$, $\Delta_d = 16.8$ K<br>$C_W = 373.4$ K<br>$\Delta_c = 3.7$ K, $\nu = 0.61$<br>$T_C = 90.8$°C |

### B. Landau-type phenomenological approach and density functional theory calculations

Analysis of the spatial-temporal evolution of polarization in the oxygen-deficient Hf$_x$Zr$_{1-x}$O$_2$ nanoparticles can be performed in the framework of "effective" LGD model [38 - 36]. This approach incorporates the elements of the Kittel-type model [92] with polar and antipolar modes [93, 94, 95] to the Landau-type free energy with effective parameters [96] extracted either from the experiment [19] or from the first principles calculations. It was shown earlier that the Landau-Devonshire expansion of the Gibbs energy $G$ in even (2-nd, 4-th and 6-th) powers of the polar order parameter ($P_3$) and the antipolar order parameter ($A_3$) after partial minimization, $\frac{\partial G}{\partial A_3} = 0$, can be transformed to the expansion in even powers (2-nd, 4-th, 6-th and 8-th) of the polarization $P_3$ [97]. The bulk density of the LGD potential, $g_{LGD}$, which includes the Landau-Devonshire expansion in even powers of the polarization $P_3$ and the Ginsburg gradient energy has the form:

$$g_{LGD} = \frac{\alpha(T)}{2} P_3^2 + \frac{\beta}{4} P_3^4 + \frac{\gamma}{6} P_3^6 + \frac{\delta}{8} P_3^8 + g_{33ij} \frac{\partial P_3}{\partial x_i} \frac{\partial P_3}{\partial x_j}. \qquad (7)$$

The temperature dependence of the coefficient $\alpha(T)$ obeys the Barrett-type law [98], $\alpha(T) = \alpha_T T_q \left( \coth \frac{T_q}{T} - \coth \frac{T_q}{T_C} \right)$, where $T_C$ is the Curie temperature and $T_q$ is the quantum vibration temperature. At low temperatures $\alpha \to \alpha_T T_q \left( 1 - \coth \frac{T_q}{T_C} \right)$. The coefficients $\beta$, $\gamma$, and $\delta$ are regarded as temperature independent. Tensor $g_{ijkl}$ is tensor of polarization gradient energy.

To determine the coefficients $\alpha$, $\beta$, $\gamma$, and $\delta$ in Eq.(7) from the first principles, we performed the density functional theory (DFT) calculations of the $Hf_xZr_{1-x}O_2$ structures using the full-potential all-electron local-orbital (FPLO) code [99]. We have used the default FPLO basis. The structural parameters of 12-atom cells are optimized with using local density approximation (LDA), a $6\times6\times6$ $k$-point mesh for Brillouin zone sampling, an energy convergence threshold of $10^{-8}$ Ry, and a force convergence threshold of $10^{-6}$ Ry/Bohr. Both lattice constants and ionic positions are fully relaxed.

The calculated equilibrium lattice parameters $a$, $b$, and $c$ were compared with experimental results [100, 101, 102] (see Ref. [39] for details and **Table S3** [41]). In agreement with earlier works, the crystal structure with a monoclinic symmetry has the lowest ground state energy in the bulk $HfO_2$, $Hf_{0.5}Zr_{0.5}O_2$ (HZO) and $ZrO_2$. However, the difference $\Delta U$ between the energy $G_f$ of the ferroelectric o-phase $Pca2_1$ (f-phase) and the energy $G_m$ of the m-phase $P2_1/c$, calculated using the FPLO basis in Ref.[39], turned out to be smaller than those calculated by other DFT methods. From **Table S3** [41], the difference between the m-HZO and f-HZO (~7.8 meV/f.u.) is significantly smaller than the difference between the m-$HfO_2$ and f-$HfO_2$ (~11.6 meV/f.u.). Thus, the addition of Zr may stabilize the f-phase. The dependence of the f-HZO total energy on the displacements of the four oxygen atoms (O1) in z-direction from their equilibrium positions in the f-phase with the minimal energy are listed in **Table S4** [41] and shown in **Fig. 10**.

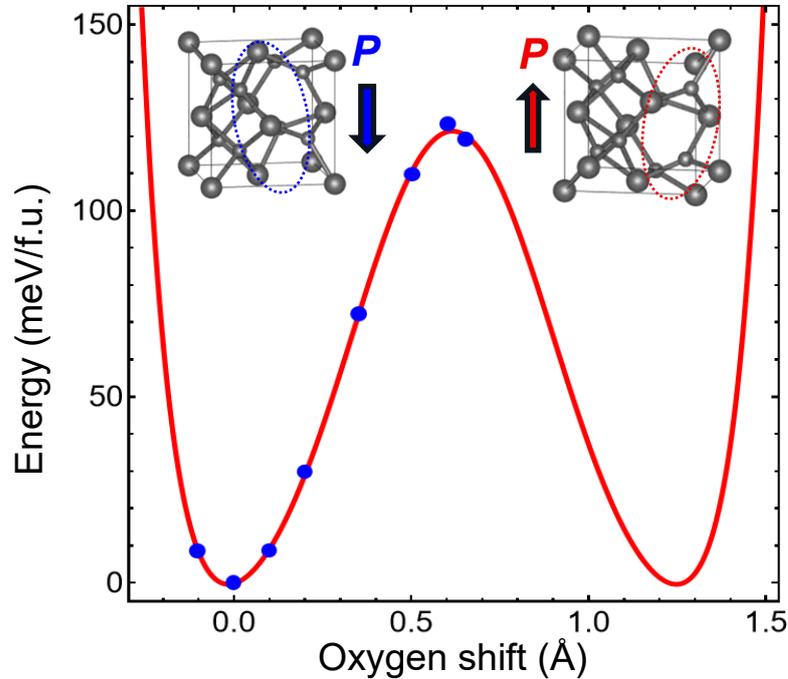

**FIGURE 10**. The dependence of the f-$Hf_{0.5}Zr_{0.5}O_2$ total energy on the displacements of the four oxygen atoms (O1) in z-direction from their equilibrium positions in the f-phase with the minimal energy. Symbols are data

from **Table S4**, solid curve is the reconstructed Landau-Devonshire potential with $\alpha$, $\beta$, $\gamma$, and $\delta$ (see **Table IV**).

**Table IV**. The values of LGD expansion coefficients

| parameter | Atomic units | SI units |
|---|---|---|
| $\alpha$ | 1712.8 **meV/(f.u. Å$^2$)** | $-1.407 \cdot 10^9 \, J \, m/C^2$ |
| $\beta$ | 9667.8 **meV/(f.u. Å$^4$)** | $5.358 \cdot 10^9 \, J \, m^5/C^4$ |
| $\gamma$ | -21879 **meV/(f.u. Å$^6$)** | $-8.181 \cdot 10^9 \, J \, m^9/C^4$ |
| $\delta$ | 21079 **meV/(f.u. Å$^8$)** | $5.318 \cdot 10^9 \, J \, m^{13}/C^4$ |

The parameters of the "classical" LGD approach, such as the Curie temperature $T_C$, the Curie-Weiss constant $C_W$, expansion coefficients $\alpha(T)$, $\beta$, $\gamma$ and $\delta$, listed in **Table IV**, are considered to characterize a bulk ferroelectric material, being material parameters per se. These parameters are typically determined from the fitting of experimentally measured temperature dependences of spontaneous polarization and dielectric permittivity by the dependences calculated from the LGD free energy (e.g., from Eq.(7)). The LGD parameters could vary slightly from sample to sample due to the deviations from stoichiometry, uncontrollable impurities and/or defects, but a total reproducibility of the results is an established fact for ordered ferroelectrics. For solid solutions (e.g., for the lead zirconate-titanate Pb$_x$Zr$_{1-x}$TiO$_3$), the LGD parameters are continuous and often monotonic functions of the solid solution composition "x".

For nanoscale ferroelectrics (e.g., thin films and nanoparticles), layered and/or for disordered ferroelectric systems (e.g., relaxor ferroelectrics, sliding ferroelectric and ferroelectric with diffuse phase transitions) the situation is completely different. As a matter of fact, the applicability of the classical LGD approach to such systems is under debate and requires significant modification, because it is impossible to separate the influence of extrinsic factors and size effects (e.g., surfaces, interfaces, substrates, shells and environment) and intrinsic inhomogeneity (e.g., layered structure, flexoelectric effect, compositional disorder and induced flexo-chemical strains) on their physical properties. For several important cases, including Hf$_x$Zr$_{1-x}$O$_2$ nanoparticles [36, 38] and thin films [96], the applicability of the LGD approach is based on the "renormalized" free energy, in which the "effective" LGD parameters, which depend on the size, surface characteristics and environment, are introduced.

In accordance with Refs. [36, 38], the effective coefficient $\alpha(T, x)$ of spherical Hf$_x$Zr$_{1-x}$O$_2$ nanoparticles covered with the shell enriched by oxygen vacancies is "renormalized" by the chemical pressure (analog of surface tension) induced by the vacancies [103, 104], depolarization field contribution and polarization gradient energy:

$$\alpha_R(T,R,x) = \alpha(T,x) - Q\frac{\mu_s}{R} + \frac{\varepsilon_0^{-1}}{\varepsilon_b + 2\varepsilon_e + (R/\Lambda_s)} + \frac{2g}{R\Lambda_P + R^2/4}. \qquad (8)$$

Here $Q = Q_{11} + 2Q_{12}$ is the combination of electrostriction coefficients $Q_{ij}$, $\mu_s$ is the coefficient of intrinsic surface stress [105], $R$ is the particle radius, $\varepsilon_0$ is a universal dielectric constant, $\varepsilon_b$ is the static background permittivity of the nanoparticles, $\varepsilon_e$ is the relative permittivity of the media surrounding the particles, $\Lambda_s$ is the effective screening length produced by the ions and/or oxygen vacancies adsorbed by the surface of the nanoparticle, $g = \frac{1}{2}(g_{3311} + g_{3311})$ is the combination of the polarization gradient energy tensor $g_{ijkl}$, and $\Lambda_P$ is the polarization extrapolation length [36].

The ferroelectric-like state may be stable in the Hf$_x$Zr$_{1-x}$O$_2$ nanoparticle when its Gibbs free energy $G_f$ is smaller than the bulk energy of the monoclinic phase $G_m$. In accordance with the DFT results listed in **Table S3** [41], the energy difference $\Delta U = G_f - G_m$ is equal to 7.8 meV/f.u. for the bulk Hf$_{0.5}$Zr$_{0.5}$O$_2$. From Eqs.(7)-(8), the condition of the f-phase stability is

$$\left(\frac{\varepsilon_0^{-1}}{\varepsilon_b + 2\varepsilon_e + (R/\Lambda_s)} + \frac{2g}{R\Lambda_p + R^2/4} - Q\frac{\mu_s}{R}\right)\frac{P_S^2}{2} < -\Delta U. \qquad (9)$$

Here $P_S \approx 75$ µC/cm$^2$ is the hypothetic spontaneous polarization of the Hf$_{0.5}$Zr$_{0.5}$O$_2$ calculated from the DFT using parameters listed in **Table IV**. The condition (9) determines the range of the nanoparticle radii for which the f-phase could be stable. Assuming that the contribution of the polarization gradient energy is much smaller than the contribution of the depolarization field energy, namely $\frac{2g}{R\Lambda_P + R^2/4} \ll \frac{\varepsilon_0^{-1}}{\varepsilon_b + 2\varepsilon_e + (R/\Lambda_s)}$, that is a good approximation for a large $\Lambda_P$ and a small $g$, it is easy to derive an approximate algebraic equation for the critical radius determination:

$$\frac{\varepsilon_0^{-1}}{\varepsilon_b + 2\varepsilon_e + (R_{cr}/\Lambda_s)} - Q\frac{\mu_s}{R_{cr}} + \frac{2\Delta U}{P_S^2} = 0. \qquad (10)$$

Assuming that $(R_{cr}/\Lambda_s) \gg \varepsilon_b + 2\varepsilon_e$, which corresponds to the case of high ionic screening by larger nanoparticles, we obtain that

$$R_{cr} \approx \frac{P_S^2}{2\Delta U}\left(\mu_s Q - \frac{\Lambda_s}{\varepsilon_0}\right). \qquad (11a)$$

The nanoparticles may become ferroelectric for $R < R_{cr}$, and the critical radius exists at $\mu_s Q > \frac{\Lambda_s}{\varepsilon_0}$. For estimates we use the parameters $P_S \approx 75$ µC/cm$^2$, $\Delta U = 7.8$ meV/(5 Å)$^3$, $Q \cong 0.2$ m$^4$/C$^2$, $\mu_s \cong 2.5$ N/m [106, 107] and $\varepsilon_0 = 8.85 \cdot 10^{-12}$ F/m, which gives that the condition $\Lambda_s < \varepsilon_0 \mu_s Q$ is valid for $\Lambda_s \leq 0.045$ Å. Despite the effective screening length is much smaller than the lattice constant (5 Å), its value may be quite realistic [108], because the effective screening length $\Lambda_s$ is related with the real charge-surface separation $h$ as $\Lambda_s \cong h/\varepsilon_{eff}$ [36]. Since the static dielectric permittivity $\varepsilon_{eff}$ is higher than $10^2 - 10^3$ and reaches $10^4$ in maximum (see the green curve in **Fig. 7**, left), the separation $h$ is higher than (4.5 – 45) Å, being about 1 – 10 lattice constants. The upper estimate of $R_{cr}$ is $R_{cr} < \frac{P_S^2}{2\Delta U}\mu_s Q$, which gives about 15 nm.

Assuming that $(R_{cr}/\Lambda_s) \ll \varepsilon_b + 2\varepsilon_e$, which corresponds to the case of weak ionic screening by smaller nanoparticles, we obtain that

$$R_{cr} \approx \mu_s Q \left(\frac{\varepsilon_0^{-1}}{\varepsilon_b+2\varepsilon_e} + \frac{2\Delta U}{P_S^2}\right)^{-1}. \tag{11b}$$

From Eq.(11b) the critical radius is about 0.1 Å that is much smaller than the lattice constant and thus has no physical sense.

Thus, we can conclude that if the concentration of oxygen vacancies is relatively small ($\Lambda_s$ >0.1 Å), the long-range ferroelectric order cannot exist in Hf$_x$Zr$_{1-x}$O$_2$ nanoparticles due to the insufficient screening of the emerging spontaneous polarization. At large concentrations of the surface ions and/or charged vacancies the synergy of high screening ($\Lambda_s \leq 0.05$ Å) and chemical pressure (large term $-Q\frac{\mu_s}{R}$) supports and stabilizes the long-range ferroelectric order in Hf$_x$Zr$_{1-x}$O$_2$ nanoparticles. Due to the weaking of size effects with increase in $R$, the imbalance of the negative surface tension ($-Q\frac{\mu_s}{R}$) and positive depolarization field energy ($\frac{\varepsilon_0^{-1}}{\varepsilon_b+2\varepsilon_e+(R/\Lambda_s)}$) contributions arises. In result the ferroelectric Hf$_x$Zr$_{1-x}$O$_2$ nanoparticles become paraelectric with an increase in their size above $2R_{cr} \cong 30$ nm, and then dielectric with its further increase [36]. A possible stability of the f-phase for $R < R_{cr}$ is the synergy of ferro-ionic coupling and size effects.

We also should underline that the observed maximum of dielectric permittivity, which may be related to the diffuse ferroelectric-like phase transition, is possibly complicated by the significant contribution of IBLC effects. Due to these reasons, the fitting parameters listed in **Table III**, which describe the dielectric response of the oxygen-deficient Hf$_x$Zr$_{1-x}$O$_2$ nanopowders, do not have a direct relation to the LGD parameters of hypothetical bulk crystalline ferroelectric Hf$_{0.5}$Zr$_{0.5}$O$_2$ listed in **Table IV**, which are determined by the DFT calculations. Since the bulk ferroelectric Hf$_{0.5}$Zr$_{0.5}$O$_2$ does not exist in nature and the nonpolar monoclinic bulk Hf$_{0.5}$Zr$_{0.5}$O$_2$ exists instead, the parameters from **Tables III** and **IV** cannot be directly linked, as they serve for completely different purposes.

The link between the DFT results and "effective" LGD parameters is given by Eq.(8), which shows how the surface and size effects renormalizes the parameter $\alpha(T)$ calculated by the DFT. Using the renormalized parameter $\alpha_R(T,R,x)$ given by Eq.(8) and the energy difference $\Delta U = G_f - G_m$ (calculated from DFT), we formulate the conditions (9)-(10), which allow us to estimate the critical radius $R_{cr}$ of Hf$_x$Zr$_{1-x}$O$_2$ nanoparticles, below which they may become ferroelectric. The value of $R_{cr}$ given by Eq.(11) well agrees with experimental observation of colossal dielectric response and o-phases in small (size less than 30 nm) Hf$_x$Zr$_{1-x}$O$_2$ nanoparticles and their absence in larger particles.

# V. DISCUSSION AND CONCLUSIONS

We analyze the dielectric response and conductivity of small (5 – 10 nm) oxygen-deficient $Hf_xZr_{1-x}O_2$ (x = 1 – 0.4) nanoparticles prepared by the solid-state organonitrate synthesis, where the dominance of the o-phases was proved by the X-ray diffraction. We reveal that the temperature dependencies of the dielectric permittivity and resistivity of the $Hf_xZr_{1-x}O_2$ nanopowders pressed at 5 MPa are almost mirror-like reflected in respect to each other, which means that all their features, such as position and shape of maxima, plateau, minima and inflexions, almost coincide after the mirror reflection in respect to the temperature axis. The correlations in behavior of the resistivity and dielectric permittivity are well-described in the Heywang model [40] for semiconductor ferroelectrics applied in combination with the variable range hopping conduction model.

The effective dielectric permittivity of the $Hf_xZr_{1-x}O_2$ nanopowders has a pronounced maximum at 311 – 361 K (38 – 88°C) for Hf content x = 1 – 0.4, the height of which decreases strongly and monotonically with an increase in frequency. The maximal dielectric permittivity increases from $1.5 \cdot 10^3$ for x = 1 to $1.3 \cdot 10^5$ for x = 0.4 at low frequencies (~4 Hz); being much smaller at high frequencies, namely changing from 7 (for x = 1) to 20 (for x = 0.4) at 500 kHz.

As a rule, the colossal effective dielectric permittivity ($\sim 10^4 - 10^6$) of ceramics and composites with ferroelectric nanograins is related with the IBLC and/or SBLC effects, and/or with polaron activation and hopping [61 - 65]. The colossal values of their dielectric permittivity can be accompanied by an increase in the conduction losses [63]. A significant frequency dispersion of the features of dielectric permittivity and losses is inherent to the IBLC, SBLC and polaronic effects. Since the frequency dispersion of the dielectric permittivity maximum and resistivity minimum is absent in the studied $Hf_xZr_{1-x}O_2$ nanopowders, it allows us to assume a signification contribution of the expected ferroelectric-like phase transition to the dielectric response. Notably that the temperature of the possible transition increases monotonically with increase in Zr content, meanwhile the shape and width of the dielectric permittivity maximum changes in a more complex way.

Notably, we do not observe any frequency dispersion of the dielectric permittivity maximum position in the $Hf_xZr_{1-x}O_2$ nanopowders inherent to relaxor ferroelectrics and dipolar glasses, but the shape and width of the maximum changes strongly with increase in frequency. At the same time, the permittivity maximum is wide or plateau-like, especially for $Hf_{0.6}Zr_{0.4}O_2$ nanopowders entire the frequency range, which may indicate that we observe a diffuse ferroelectric-like phase transition, which classification is much wider in comparison with relaxor ferroelectrics.

The ferroelectric-like behavior of the small (radius less than 15 nm) oxygen-deficient $Hf_xZr_{1-x}O_2$ nanoparticles is expected from the density functional theory calculations and Landau-Ginzburg-Devonshire approach, which reveal that the oxygen vacancies create chemical pressure inside the particle and provide effective ionic screening at its surface. The problem of direct evidence of ferroelectricity in small $Hf_xZr_{1-x}O_2$ nanoparticles may be solved by preparation of dense ceramics, where the fraction of nanoparticles exceeds 90 %. However, conventional hot-pressed ceramics are not suitable for this purpose, because the sintering temperature of hafnia-zirconia is rather high (above 900°C), which inevitably lead to the agglomeration of ultra-small nanoparticles to the larger grains with the average size larger than 30-50 nm. Such grains cannot be ferroelectric according to theoretical calculations of $Hf_xZr_{1-x}O_2$ nanoparticles polar properties [36] and experimental results showing the ferroelectricity disappearance in $Hf_xZr_{1-x}O_2$ films of thickness above 30 nm [1].

Obtained results may be useful for developing next generation of silicon-compatible ferroelectric nanomaterials. Indeed, the ferroelectrics of next generation, such as $Hf_xZr_{1-x}O_2$, $Al_xSc_{1-x}N$ and $Zn_xMg_{1-x}O$, are regarded Si-compatible, because they either demonstrate a significant resistance to oxygen loss or have no oxygen at all, as well as they do not require crystallization at high temperatures (see e.g., Refs.[1-5] and refs therein). However, the coercive fields of these ferroelectrics are relatively high [1-5]. The coercive field reduction can be achieved due to the appropriate doping [109, 110, 111], proximity effects [112, 113] and/or finite size effects [36, 96]. It is important to maintain the Si-compatibility together with the reduction of coercive field. For instance, Jiang et al. [114] achieved an ultra-low-voltage switching of ferroelectric polarization in ultrathin $BaTiO_3$ films, which direct integration with Si-substrate requires a $SrTiO_3$-buffer layer. The Si-compatibility of the studied $Hf_xZr_{1-x}O_2$ nanoparticles can be reached by their "cold" sputtering on the appropriate silicon of silicon-oxide substrate; at that the low deposition temperature (< 300°C) is required to prevent their agglomeration. Not less important seems to explore the possibility of their coercive field reduction by size effects. In this sense, small sizes of the nanoparticles can reduce the coercive field significantly due to the depolarization field emerging in the small particles under incomplete screening conditions.

**Authors' contribution**

The research idea belongs to A.N.M. and V.V.V. O.S.P., V.V.V. and V.N.P. performed electrophysical measurements and interpreted results. O.V.L. and V.N.P. sintered the $Hf_xZr_{1–x}O_2$ nanopowders and characterized them by TDA. S.E.I. sintered the $BaTiO_3$ nanopowders. I.V.K. performed DFT calculations. L.D. performed TEM studies. M.V.K. performed XRD studies. E.A.E. and A.N.M. performed analytical calculations and fitting the experimental results. A.N.M. and V.V.V. wrote the manuscript draft. All co-authors discussed and analyzed the results, and corresponding authors made all improvements in the manuscript.

**Acknowledgments**

Authors gratefully acknowledge useful remarks, suggestions and discussions with Prof. Mikhail P. Trubitsyn (DNU), Dr. Yuriy O. Zagorodniy, Dr. Roman A. Kuzian (IPMS NASU) and Dr. Nicholas

V. Morozovsky (IoP NASU). Also, authors are very grateful to the Reviewers for the constructive suggestions and stimulating discussions. The work of O.S.P. and A.N.M. are funded by the National Research Foundation of Ukraine (grant N 2023.03/0132 "Manyfold-degenerated metastable states of spontaneous polarization in nanoferroics: theory, experiment and perspectives for digital nanoelectronics"). The work of E.A.E., I.V.K., L.P.Y. and A.O.D. are funded by the National Research Foundation of Ukraine (grant N 2023.03/0127 "Silicon-compatible ferroelectric nanocomposites for electronics and sensors"). V.V.V. and V.N.P. acknowledge the Target Program of the National Academy of Sciences of Ukraine, Project No. 5.8/25-П "Energy-saving and environmentally friendly nanoscale ferroics for the development of sensorics, nanoelectronics and spintronics". A.N.M. also acknowledges the support (materials characterization) from the Horizon Europe Framework Programme (HORIZON-TMA-MSCA-SE), project № 101131229, Piezoelectricity in 2D-materials: materials, modeling, and applications (PIEZO 2D). Preparation and characterization of some materials is sponsored by the NATO Science for Peace and Security Programme under grant SPS G5980 "FRAPCOM". L.D. acknowledges the grant SSF UKR24-0007 from the Swedish Foundation for Strategic Research for conducting TEM research.

**Data availability**

The experimental data is available from the authors upon reasonable request. The codes for fitting are written Mathematica 14.0 and are freely available at [115].

# Supplementary Materials
# "Colossal dielectric response of Hf$_x$Zr$_{1-x}$O$_2$ nanoparticles"

**APPENDIX S1. XRD analysis of Hf$_x$Zr$_{1-x}$O$_2$ nanopowders, their synthesis and samples preparation details**

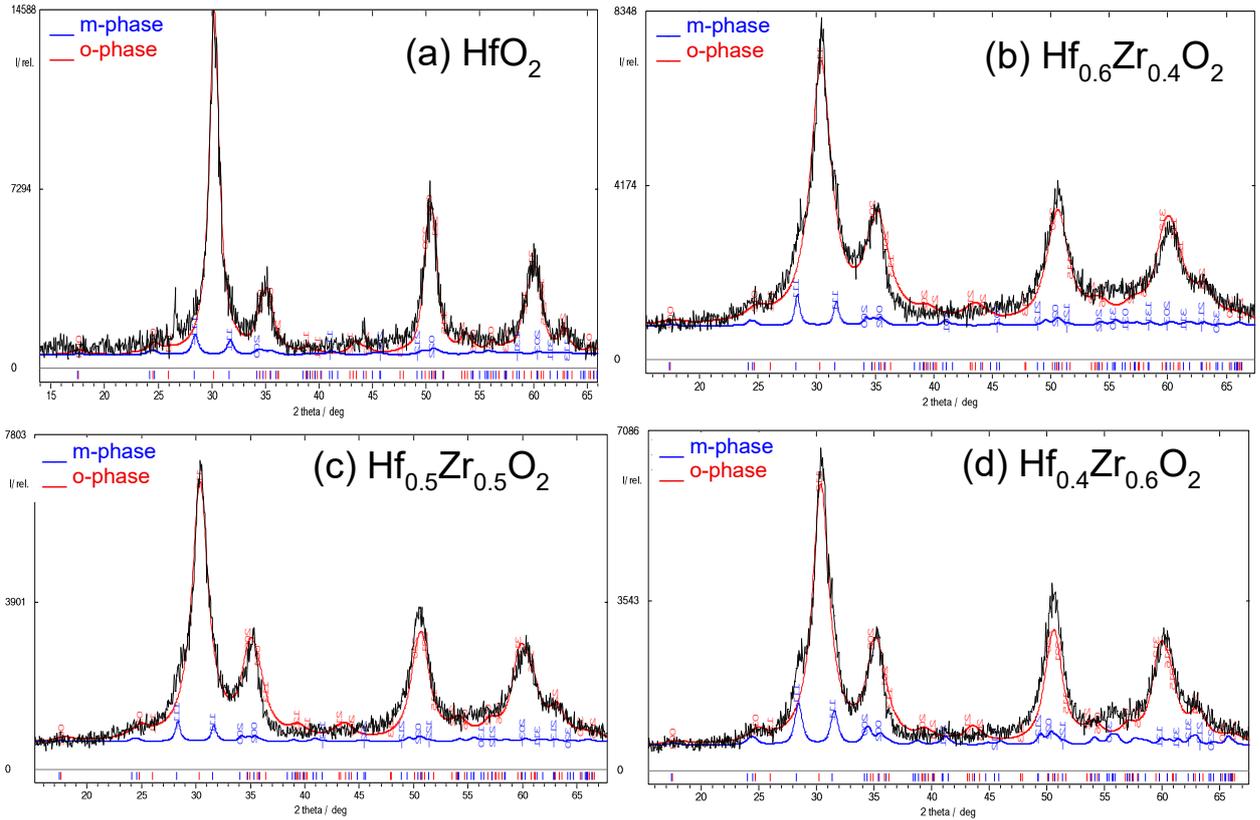

**FIGURE S1.** XRD spectra of the Hf$_x$Zr$_{1-x}$O$_2$ nanopowders.

Synthesis details. The studied BaTiO$_3$ nanopowders were synthesized using the non-isothermal decomposition method of high-purity barium titanyl oxalate Ba(TiO)(C$_2$O$_4$)$_2$ produced by the Degussa Electronic Corporation (Netherlands). The non-isothermal decomposition of barium titanyl oxalate was carried out at temperatures from 650 to 700°C in a specially designed apparatus. The apparatus allows thermal decomposition under various gaseous atmospheres, with heating rates ranging from 5°C/h to 1500°C/h, up to a temperature 900°C, and under conditions of intensive mixing.

Samples preparation details. We used a 3-contact circuit to determine the contribution of electric contacts to the measured dielectric response (see **Fig.S2(a)-(b)**). It appeared that the voltage drop across the contact does not exceed in the worst-case maximum about 2-4 percent against the whole voltage drop across the sample (see **Fig.S2(c)**). Different electrode materials, such as simple Cu foil or Cu foil covered by In, did not show noticeable difference.

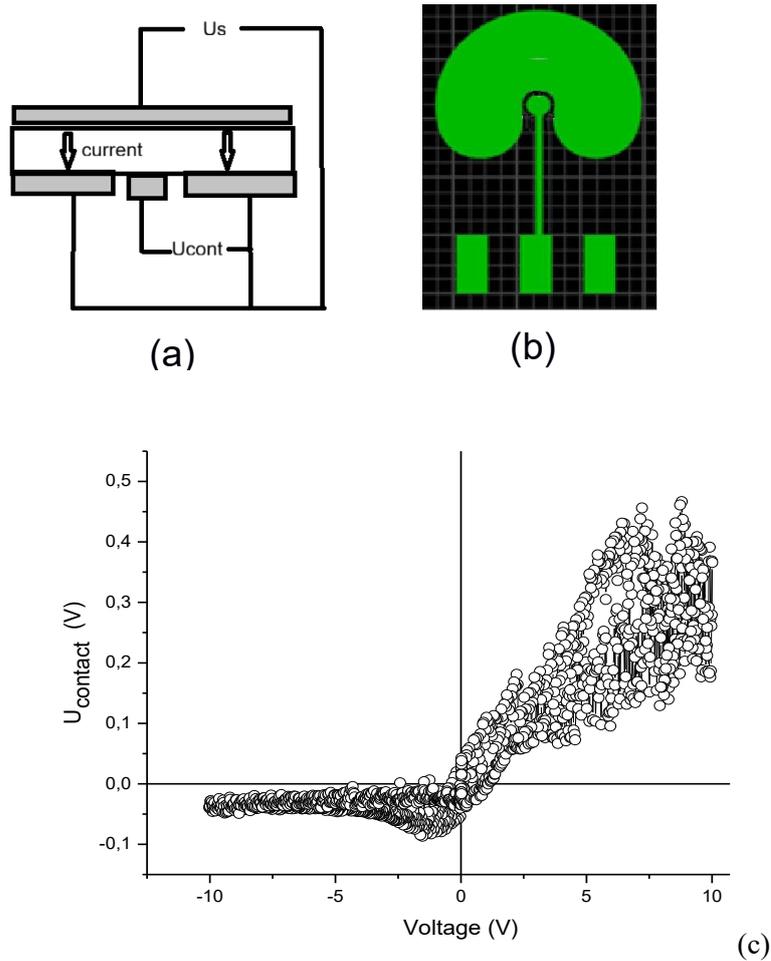

**Figure S2**. (**a**) Three-contact electric circuit to investigate the voltage drop (Ucont) across the current contact. An empty rectangle is the sample; grey rectangles are the electric contacts. At the bottom sample side, one makes 2 contacts: a large current contact under study and small (central) contact to measure the voltage drop across the current contact. (**b**) Image of the bottom plate with 2 contacts used to implement the 3-contact circuit: a large peripheral current contact and a small central one. (**c**) Dependence of the voltage drop across the current contact on the voltage applied to the whole sample.

## APPENDIX S2. Temperature Measurements of Charge-Discharge Currents

We performed measurements of the temperature dependences the charge accumulated in the sample in the course of current flow in the DC regime. For this purpose, we measured and fitted current vs time dependences at a constant voltage drop across a sample (see e.g., **Fig. S3(a)-(g)**). To calculate the accumulated charge, we use the exponential fitting functions of the measured current:

$$J(t) = J_0 + \sum_{i=1}^{N} A_i \exp\left(-\frac{t}{\tau_i}\right). \tag{S1}$$

Corresponding accumulated charge is given by expression:

$$\Delta Q = \int_0^\infty dt\, (J(t) - J_0) = \sum_{i=1}^{N} A_i \tau_i. \tag{S2}$$

Time dependences of the charging current measured at fixed voltage (2V) and different temperatures $T$ are shown in **Fig. S3(a)-(g)**. Symbols are experimental results; solid and dashed curves are exponential fitting with three and two relaxational terms in Eq.(S1), respectively.

The total charge $\Delta Q$ accumulated in the HZO3 sample as a function of temperature is shown in **Fig. S3(h)**. Open and solid red symbols are obtained using the fitting function (S1) with two and three exponents, respectively. Solid and dashed curves are their interpolations. It is seen from **Fig. S3(h)**, that the electric charge accumulated by the pressed nanopowders decreases strongly above 70°C. We related the observed decrease to the disappearance of the pyroelectric-type charge in the system above 90°C.

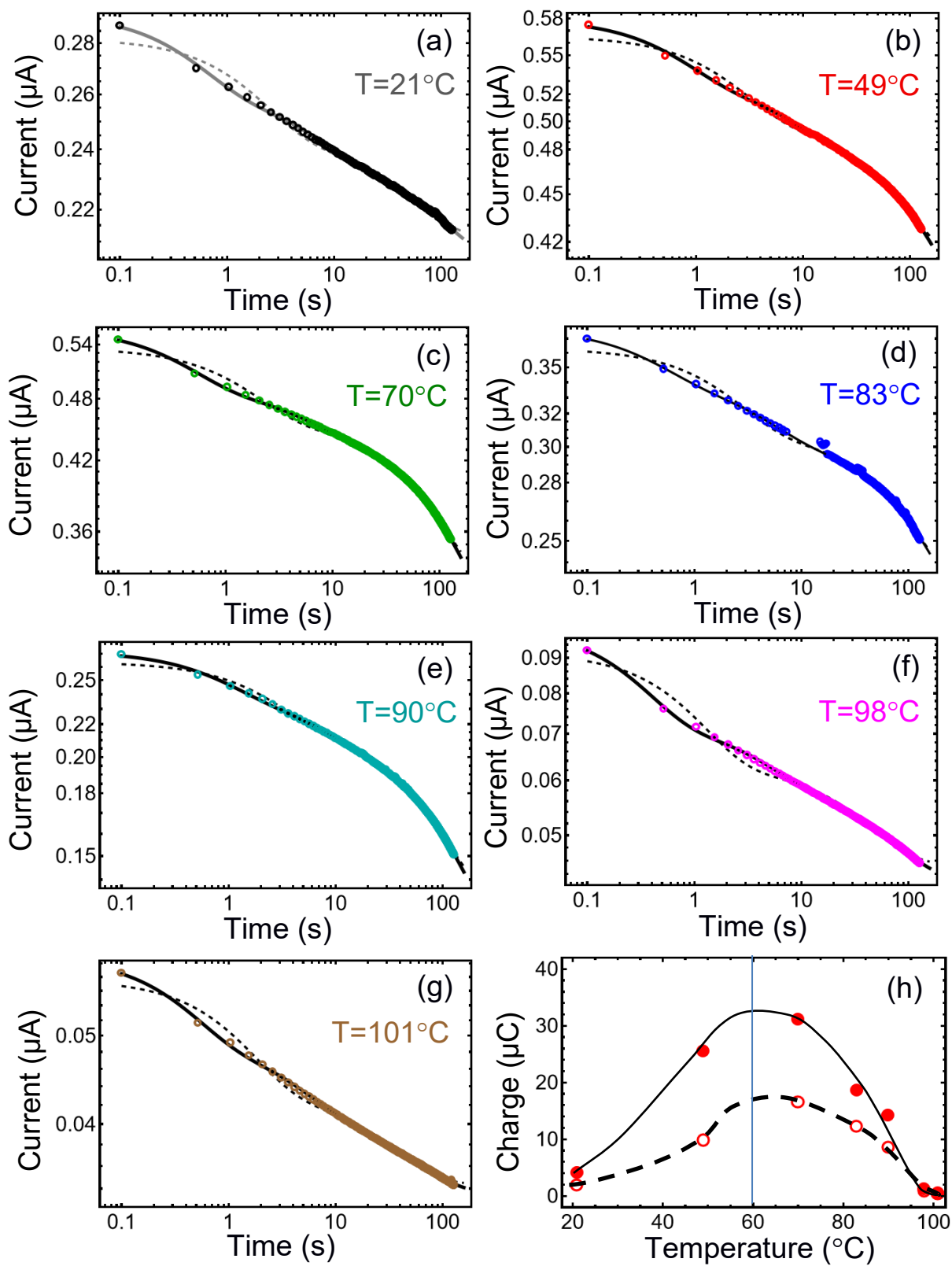

**Figure S3**. **(a-g)** Time dependence of the charging current at fixed voltage (2V) and different temperatures $T$, which are shown in the panels (a)-(g). Symbols are experimental results; solid and dashed curves are exponential fitting with three and two relaxational terms, respectively. **(h)** The total amount charge accumulated in the HZO3 sample, calculated by the integration of the varying part of the current, as a function of temperature. Open and solid red symbols are obtained using the different fitting functions with two and three exponents, respectively. Solid and dashed curves are their interpolations.

For the HZO3 sample, the accumulated charge is maximal in the temperature range 55 – 65°C, and the current is maximal near 50°C. Important, that the total discharge time (after the voltage removal) is more than 10 – 20 minutes, which may indicate the pronounced electret-like properties of the samples. It is possible to estimate the effective charge density, using the integral area of contacts, 0.1256 cm². This gives the maximal value 254 µC/cm².

## APPENDIX S3. Analysis of Nyquist plots

In general, the Nyquist plots for all studied samples show only one arc being close to a semicircle characteristic for the simple RC-circuit connected in parallel. However, we observed certain deviations of Nyquist plots from the semicircle, characteristic for the simple RC-circuit. Our semicircle has a squashy shape and its pronounced right "shoulder" may be related to a chaotic spatial fluctuation character of grains properties and their shells and consequent impact on the low-frequency conduction. The sample volume in such case may be simulated by a chaotic network of capacitors and resistors with conduction through it obeying, in principle, the percolation theory. Such deviations can be described in the framework of effective medium approximations, which are considered in detail in **Section IV.A**.

We also should note the following. A typical frequency dependence of the total impedance and phase angle of the samples under study is shown in **Fig. S4(a)** for the pressed Hf$_{0.5}$Zr$_{0.5}$O$_2$ nanopowder. The dependencies look typical for a single-component RC circuit shown in **Fig. S4(b)**.

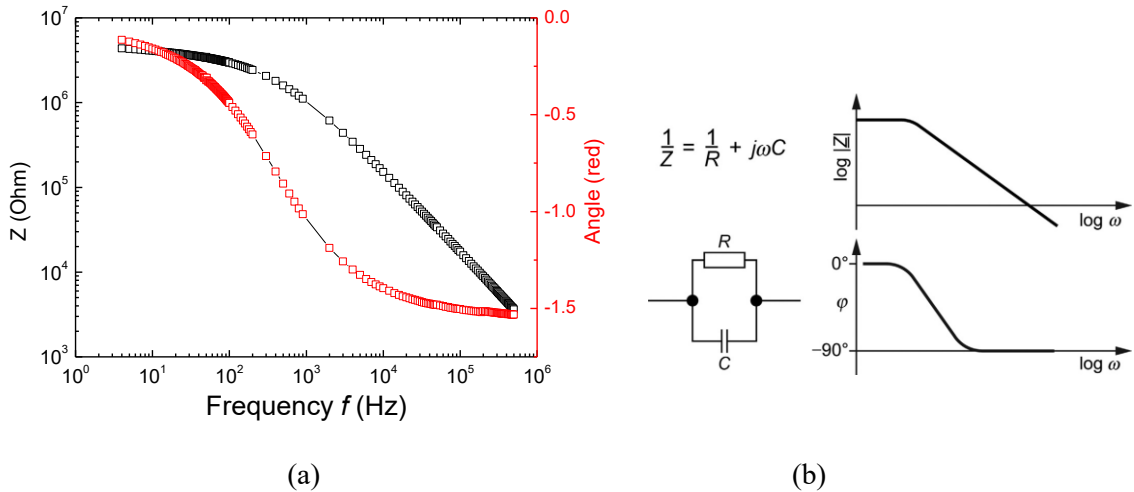

(a)           (b)

**Figure S4**. **(a)** The frequency dependence of the total impedance $|Z^*|$ and phase angle $\varphi$ measured for the pressed Hf$_{0.5}$Zr$_{0.5}$O$_2$ nanopowder. **(b)** Typical dependencies for a single-component RC cell. Part (b) is reprinted from [Electrochemical Measurement Methods and Characterization on the Cell Level Fuel Cells and Hydrogen. From Fundamentals to Applied Research. 2018, Pages 175-214. https://doi.org/10.1016/B978-0-12-811459-9.00009-8]

The complex impedance $Z^*$ of the single-component RC-cell shown in **Fig. S4(b)** is

$$Z^* = \frac{R}{1+i\omega RC} \equiv Z_{Re} + Z_{Im}, \quad Z_{re} = \frac{R}{1+(\omega RC)^2}, \quad Z_{Im} = -i\frac{\omega R^2 C}{1+(\omega RC)^2}. \tag{S3}$$

Noticeable deviations of the Nyquist plots of pressed $Hf_xZr_{1-x}O_2$ nanopowders from the semicircle are observed in the low-frequency range (see e.g., **Fig. S5**). Open black rectangles in **Fig. S5** are experimental results for $Hf_{0.5}Zr_{0.5}O_2$ nanopowders. The red curve is their fitting by semicircle $Z_{Im} = \sqrt{(1.95 \cdot 10^6)^2 - (Z_{Re} - 1.95 \cdot 10^6)^2} - 5.8 \cdot 10^5$. Since the semicircle is somewhat squashy shaped it evidences on the disordered structure of the studied powders.

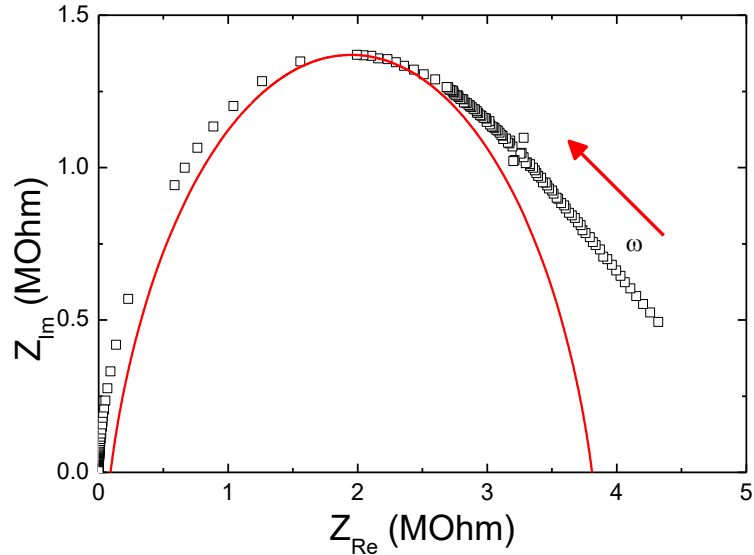

**Figure S5**. The Nyquist plot of the pressed $Hf_{0.5}Zr_{0.5}O_2$ nanopowder (open black rectangles) and its fitting by a semicircle (red curve).

From Eq.(S3) we obtain the frequency dependence of the polarizability relaxation time $\tau$, which determines the dielectric permittivity

$$\tau = RC = \frac{Z_{Im}}{\omega Z_{Re}}. \tag{S4}$$

The frequency dependence $\tau(\omega)$ is well approximated by a power law $\tau \sim \frac{1}{\sqrt{\omega}}$ (see **Fig. S5**). However, it is evident that the dependence exhibits a slight downward slope as the frequency decreases around 100 Hz, and then quickly returns to the power law. In overall, this behavior shows that ionic conductivity makes a significant contribution in the milliseconds and longer time ranges. Macroscopic ion displacement alters the field distribution within the sample.

The colossal permittivity is associated with charge accumulation in the sample, which acts as a supercapacitor at low frequencies. This is fully consistent with our assumptions and Heiwang barrier model for ferroelectrics-semiconductors [116] used in our work, because the free charges accumulate at the surface of the $Hf_xZr_{1-x}O_2$ nanoparticles due to the "net" barrier created by the uncompensated bound charges (polarization dipoles).

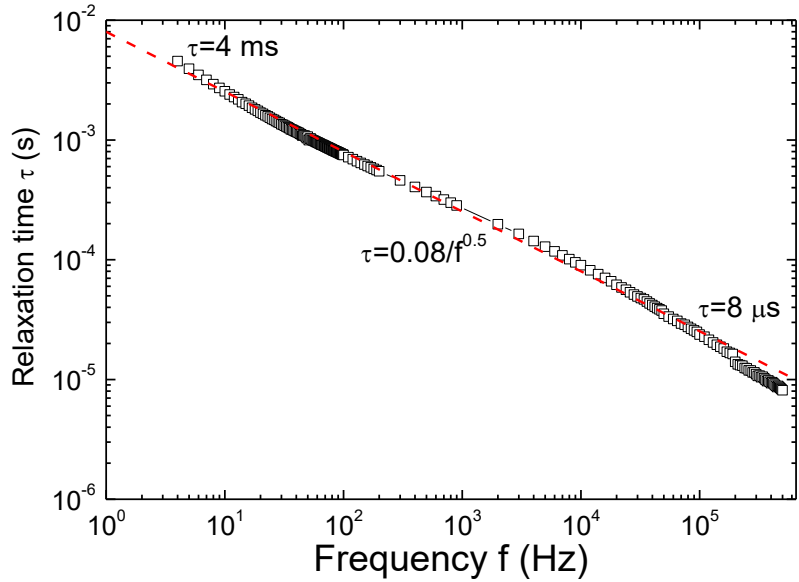

**Figure S5.** Frequency dependence of the time constant $\tau$.

### APPENDIX S4. Modelling of resistivity and effective dielectric response

The resistivity vs temperature dependence of the Hf$_{0.5}$Zr$_{0.5}$O$_2$ sample in the modified Arrhenius coordinates is shown in **Fig. S6.** Temperature dependences of the effective dielectric permittivity of the Hf$_x$Zr$_{1-x}$O$_2$ nanopowders with different Hf content are shown in **Fig. S7 – S9.**

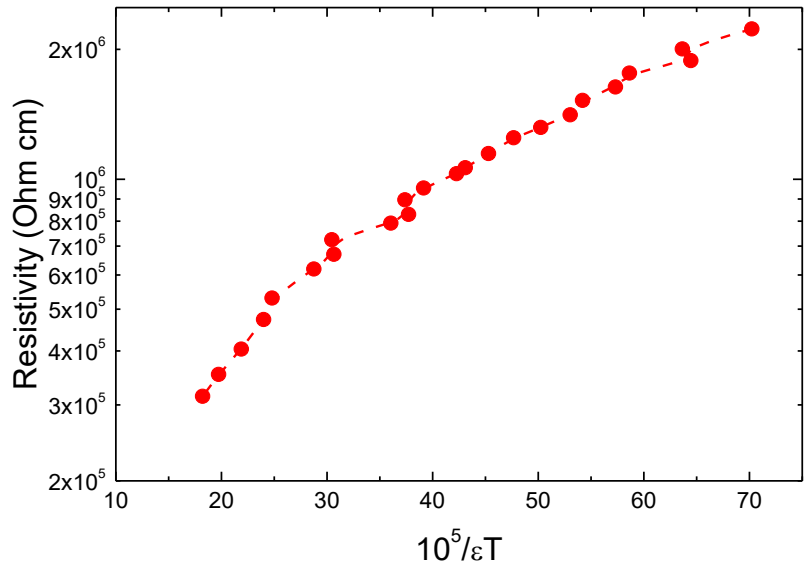

**FIGURE S6.** The resistivity vs temperature dependence of the Hf$_{0.5}$Zr$_{0.5}$O$_2$ sample in the modified Arrhenius coordinates $\ln(R)$ vs. $\frac{1}{\varepsilon T}$, where one expects the proportionality $R \sim \exp\left(\frac{A}{\varepsilon T}\right)$. The test signal frequency is 500 kHz.

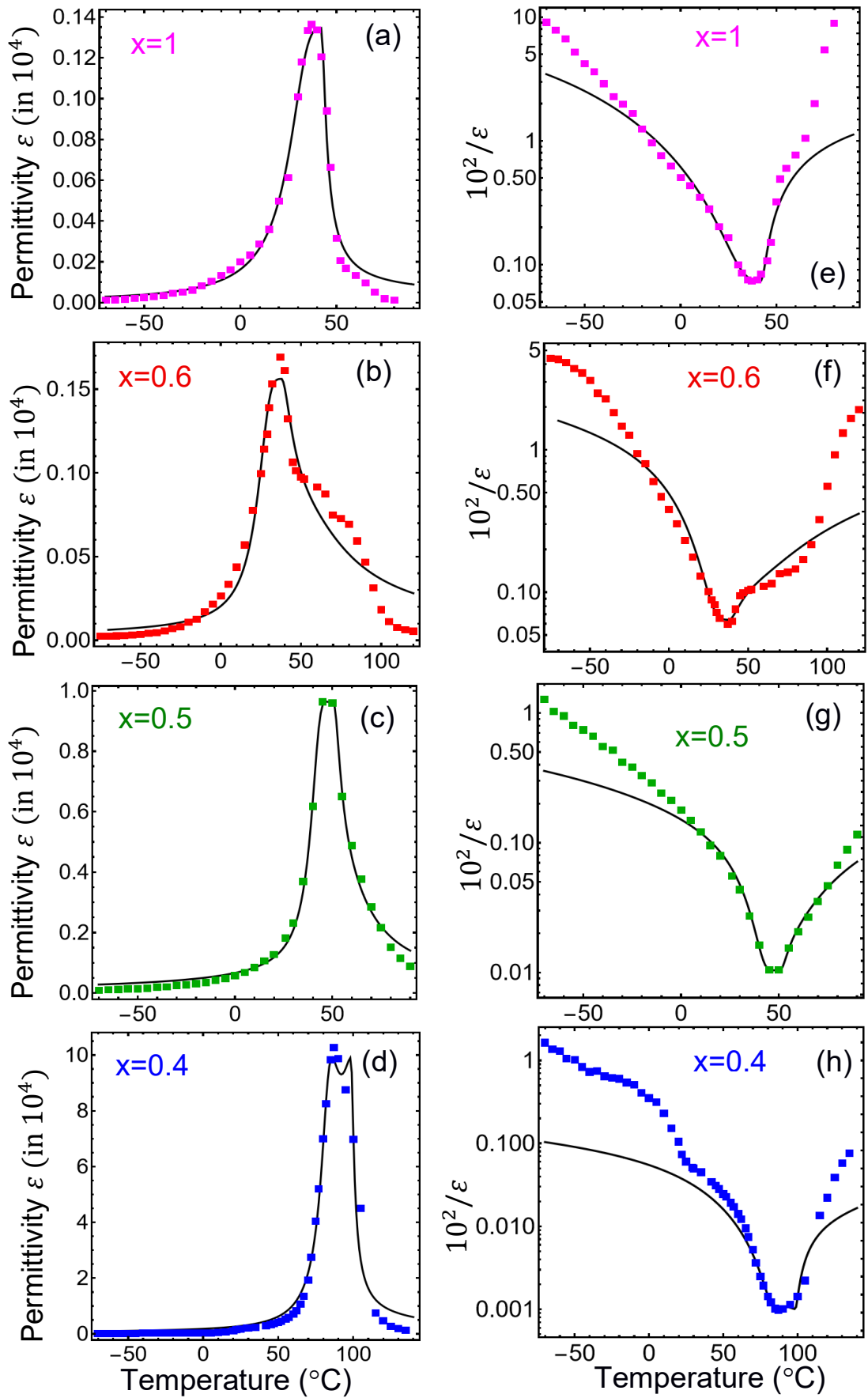

**FIGURE S7**. Temperature dependences of the effective dielectric permittivity of the $Hf_xZr_{1-x}O_2$ nanopowders with different Hf content x=1 **(a, e)**, 0.6 **(b, f)**, 0.5 **(c, g)** and 0.4 **(d, h)**. The "direct" **(a, b, c, d)** and inverse permittivity **(e, f, g, h)** are shown. Symbols are experimentally measured values at frequency 4 Hz. Solid curves are the fitting using Eq.(6a) and (6b). The fitting parameters are listed in **Table S1.**

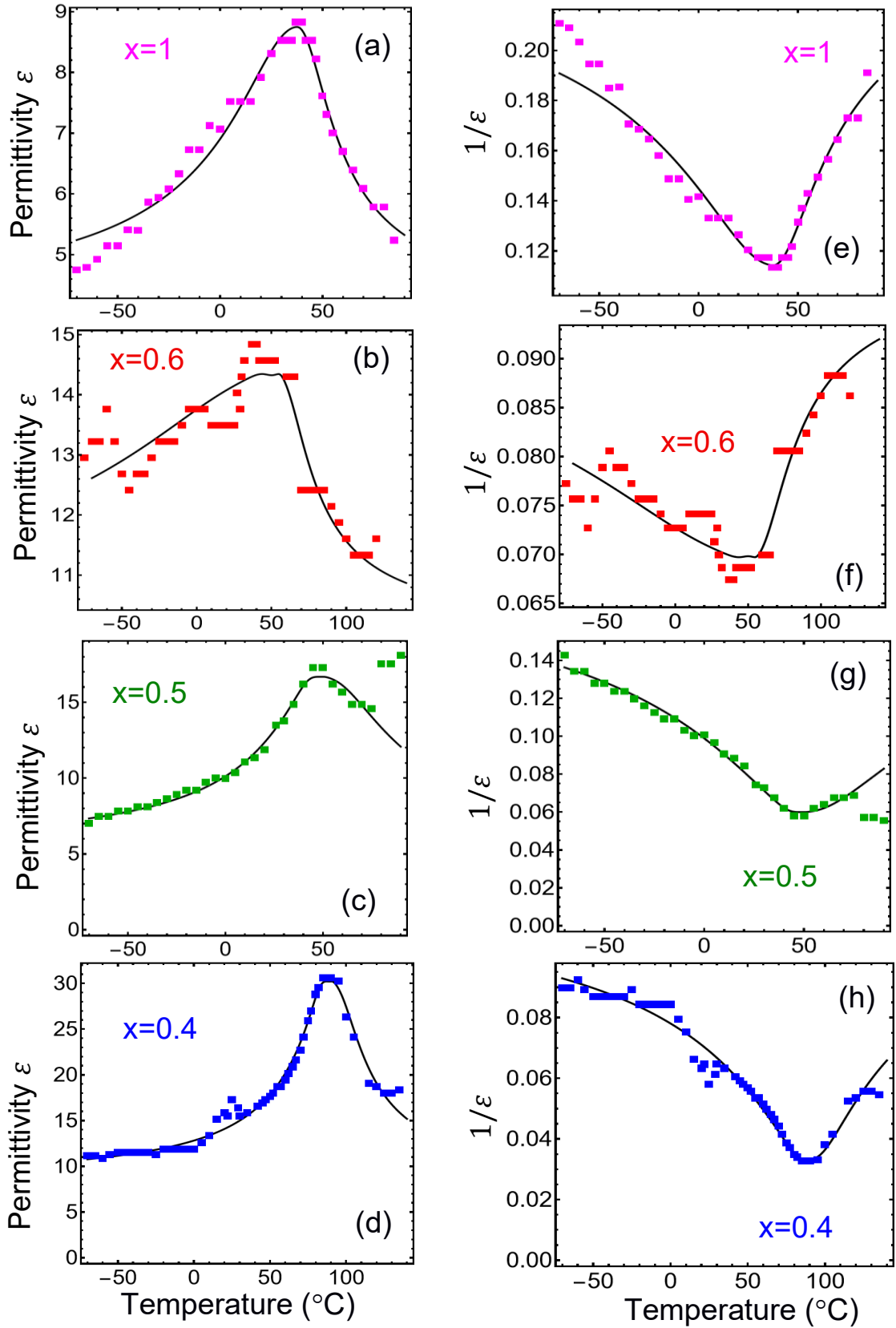

**FIGURE S8**. Temperature dependences of the effective dielectric permittivity of the $Hf_xZr_{1-x}O_2$ nanopowders with different Hf content x=1 **(a, e)**, 0.6 **(b, f)**, 0.5 **(c, g)** and 0.4 **(d, h)**. The "direct" **(a, b, c, d)** and inverse permittivity **(e, f, g, h)** are shown. Symbols are experimentally measured values at frequency 500 kHz. Solid curves are the fitting using Eq.(6a) and (6b). The fitting parameters are listed in **Table S1**.

**TABLE S1.** The values of the fitting parameters used to describe the measured dielectric permittivity

| Sample | x | Fitting using Eq.(6b), $f = 4$ Hz | Fitting using Eq.(6b), $f = 500$ kHz |
|---|---|---|---|
| HZO1 | 1 | $\varepsilon_b^0 = 4$, $\Delta_d = 4.5$ K<br>$C_W = 6080.6$ K<br>$\Delta_c = 67.3$ K, $v = 1.50$<br>$T_C = 42.0°C$ | $\varepsilon_b^0 = 4$, $\Delta_d = 15.2$ K<br>$C_W = 72.1$ K<br>$\Delta_c = 0.3$ K, $v = 0.50$<br>$T_C = 37.7°C$ |
| HZO2 | 0.6 | $\varepsilon_b^0 = 5$, $\Delta_d = 14.8$ K<br>$C_W = 23021.4$ K<br>$\Delta_c = 15.2$ K, $v = 1.57$<br>$T_C = 37.9°C$ | $\varepsilon_b^0 = 5$, $\Delta_d = 61.3$ K<br>$C_W = 558.3$ K<br>$\Delta_c = 2.8$ K, $v = 0.45$<br>$T_C = 44.3°C$ |
| HZO3 | 0.5 | $\varepsilon_b^0 = 5$, $\Delta_d = 5.77$ K<br>$C_W = 55626.5$ K<br>$\Delta_c = 8.5$ K, $v = 0.89$<br>$T_C = 50.6°C$ | $\varepsilon_b^0 = 5$, $\Delta_d = 30.03$ K<br>$C_W = 350.7$ K<br>$\Delta_c = 3.9$ K, $v = 0.74$<br>$T_C = 50.5°C$ |
| HZO4 | 0.4 | $\varepsilon_b^0 = 8$, $\Delta_d = 2.8$ K<br>$C_W = 277270.2$ K<br>$\Delta_c = 29.7$ K, $v = 0.91$<br>$T_C = 98.4°C$ | $\varepsilon_b^0 = 8$, $\Delta_d = 16.8$ K<br>$C_W = 373.4$ K<br>$\Delta_c = 3.7$ K, $v = 0.61$<br>$T_C = 90.8°C$ |

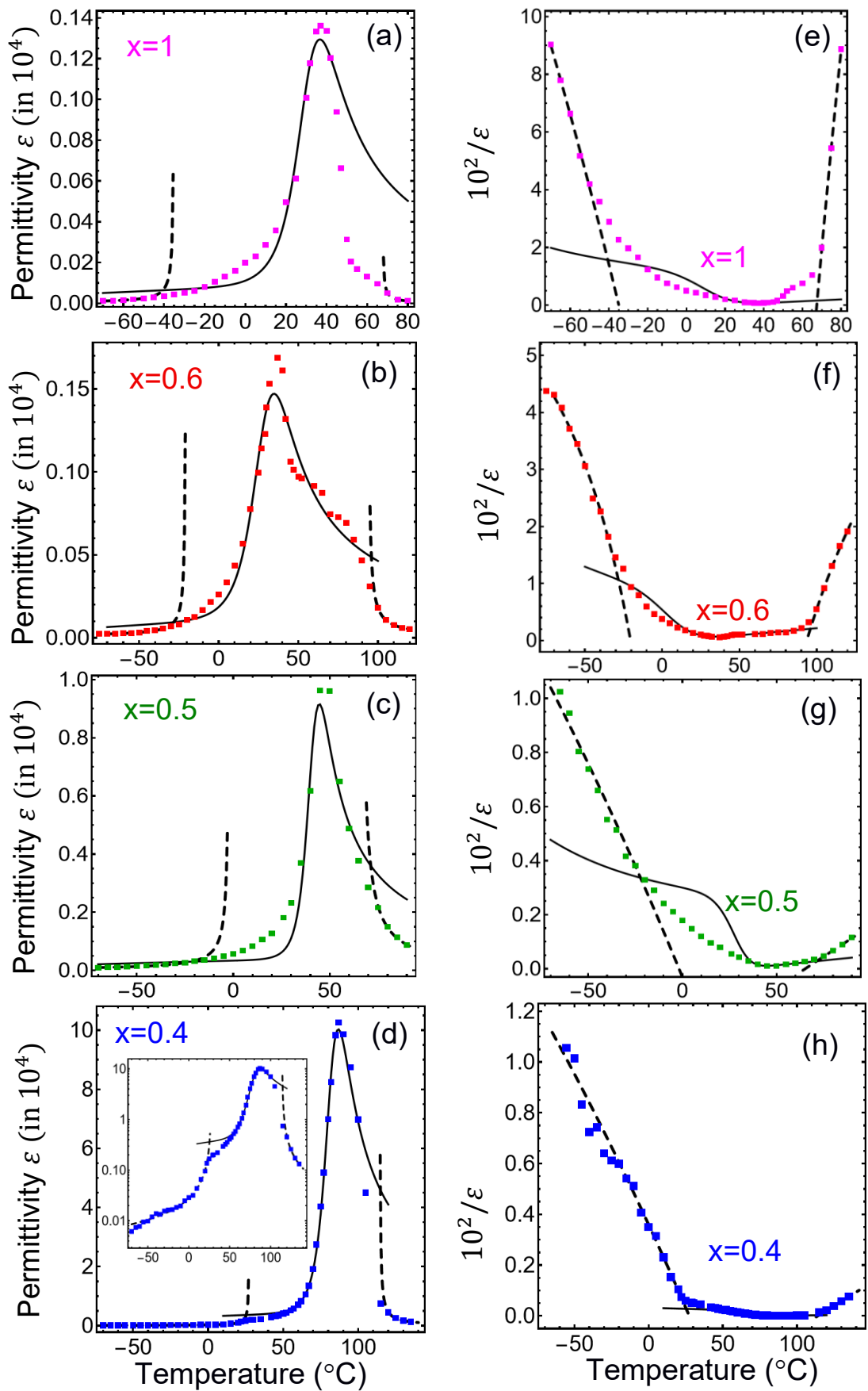

**FIGURE S9**. Temperature dependences of the effective dielectric permittivity of the $Hf_xZr_{1-x}O_2$ nanopowders with different Hf content x=1 **(a, e)**, 0.6 **(b, f)**, 0.5 **(c, g)** and 0.4 **(d, h)**. The "direct" **(a, b, c, d)** and inverse permittivity **(e, f, g, h)** are shown. Symbols are experimentally measured values with the test signal frequency

4 Hz. Solid curves are the fitting using Eq.(6a) and (6c), dashed curves correspond to the classical Curie-Weiss behavior. The fitting parameters are listed in **Table S2**.

**TABLE S2.** The values of the fitting parameters used to describe the measured dielectric permittivity

| Sample | x | Fitting using Eq.(6c), $f =4$ Hz | Curie-Weiss model $\varepsilon_{CW}(T)$ using Eq.(S1) |
|---|---|---|---|
| HZO1 | 1 | $\varepsilon_b^0 = 10$<br>$C_W = 3.02 \cdot 10^4 \, K$<br>$T_C = 219.9 \, K, \Delta_c = 10.0 \, K$ | $C_L = 355.2 \, K, T_{CL} = -34.9 \, °C, C_R = 129.9 \, K,$<br>$T_{CR} = 67.4 \, °C, \varepsilon_b^0 = 1$ |
| HZO2 | 0.6 | $\varepsilon_b^0 = 10$<br>$C_W = 3.85 \cdot 10^4 \, K$<br>$T_C = 288.1 \, K, \Delta_c = 11.7 \, K$ | $C_L = 661.2 \, K, T_{CL} = -20.5 \, °C, C_R = 1103 \, K,$<br>$T_{CR} = 93.6 \, °C, \varepsilon_b^0 = 10.0$ |
| HZO3 | 0.5 | $\varepsilon_b^0 = 15$<br>$C_W = 2.5 \cdot 10^6 \, K$<br>$T_C = 322 \, K, \Delta_c = 13 \, K$ | $C_L = 5877 \, K, T_{CL} = -1.8 \, °C, C_R =$<br>$2.16 \cdot 10^4 \, K, T_{CR} = 65.2 \, °C, \varepsilon_b^0 = 10.0$ |
| HZO4 | 0.4 | $\varepsilon_b^0 = 10$<br>$C_W = 2.01 \cdot 10^6 \, K$<br>$T_C = 344.0 \, K, \Delta_c = 7.69 \, K$ | $C_L = 7351 \, K, T_{CL} = 27.4 \, °C,$<br>$C_R = 2.74 \cdot 10^4 \, K, T_{CR} = 114.1 \, °C, \varepsilon_b^0 = 10.0$ |

$$\varepsilon_{Curie-Weiss}(T) = \begin{cases} \varepsilon_b^0(T) + \frac{C_L}{T_{CL}-T}, & T < T_{CL}, \\ \varepsilon_b^0(T) + \frac{C_R}{T-T_{CR}}, & T > T_{CR}. \end{cases} \quad (S3)$$

## APPENDIX S5. Details of the DFT calculations

**TABLE S3.** Comparison of lattice constants *a*, *b* and *c* of a 12 atomic cell and the total energy difference $\Delta U$ between the m-phase and the f-phase cells (relative to the m-phase) calculated by the DFT for HZO, HfO$_2$ and ZrO$_2$. Adapted from Ref.[O. S. Pylypchuk, I. V. Fesych, V. V. Vainberg, Y. O. Zagorodniy, V. I. Styopkin, J. M. Gudenko, I. V. Kondakova, L. P. Yurchenko, V. N. Pavlikov, A. O. Diachenko, M. M. Koptiev, M. D. Volnyanskii, V. V. Laguta, E. A. Eliseev, M. P. Trubitsyn, and A. N. Morozovska. Resistive switching and charge accumulation in Hf$_{0.5}$Zr$_{0.5}$O$_2$ nanoparticles. Journal of Physical Chemistry **29**, 31, 14299–14310 (2025); https://doi.org/10.1021/acs.jpcc.5c04140]

| Structure | $\Delta U$ (meV/f.u.) | $V$(Å$^3$) | Monoclinic angle (deg) | $a$ (Å) | $b$ (Å) | $c$ (Å) | Method |
|---|---|---|---|---|---|---|---|
| m-HZO | - | | 80.3° | 5.09 | 5.16 | 5.25 | LDA |
| f-HZO | 7.8 | 131.3 | | 5.01 | 5.21 | 5.03 | LDA |
| | - | 132.3 | | 5.01 | 5.24 | 5.04 | Exp. |
| m-HfO$_2$ | - | | 80.3° | 5.07 | 5.14 | 5.24 | LDA |
| | - | 135.8 | | 5.07 | 5.14 | 5.29 | Exp. |
| f-HfO$_2$ | 11.6 | 129.5 | | 4.99 | 5.18 | 5.01 | LDA |
| m-ZrO$_2$ | - | | 80.4° | 5.12 | 5.20 | 5.28 | LDA |
| | - | 140.3 | | 5.15 | 5.20 | 5.32 | Exp. |
| f-ZrO$_2$ | 6.6 | 133.7 | | 5.05 | 5.22 | 5.07 | LDA |
| | - | 135.5 | | 5.07 | 5.26 | 5.08 | Exp. |

**TABLE S4.** The dependence of the $Hf_{0.5}Zr_{0.5}O_2$ total energy on the displacements of the four oxygen atoms (O1) in z-direction from the positions with the minimal energy

| Shift O1 (Å) in z-direction | EE-EE$_{min}$ (meV/f.u.) |
|---|---|
| -0.1006 | 8.4 |
| 0.0 | 0.0 |
| 0.1006 | 8.55 |
| 0.2012 | 29.725 |
| 0.3521 | 72.2 |
| 0.503 | 109.65 |
| 0.6036 | 123.225 |
| 0.6539 | 118.97 |

**References**

**References**

[1]     T. Mikolajick, S. Slesazeck, H. Mulaosmanovic, M. H. Park, S. Fichtner, P. D. Lomenzo, M. Hoffmann, U. Schroeder. Next generation ferroelectric materials for semiconductor process integration and their applications, J. Appl. Phys. **129**, 100901 (2021); https://doi.org/10.1063/5.0037617

[2]     B. Mandal, A.-M. Philippe, N. Valle, E. Defay, T. Granzow, and S. Glinsek. Ferroelectric $HfO_2$–$ZrO_2$ Multilayers with Reduced Wake-Up. ACS Omega, **10** (13), 13141 (2025); https://doi.org/10.1021/acsomega.4c10603

[3]     J.Li, S.Deng, L. Ma *et al.* Enhancing ferroelectric stability: wide-range of adaptive control in epitaxial $HfO_2/ZrO_2$ superlattices. Nat. Commun. **16**, 6417 (2025); https://doi.org/10.1038/s41467-025-61758-2

[4]     J.P.B. Silva et al. Roadmap on ferroelectric hafnia and zirconia-based materials and devices, APL Mater. **11**, 089201 (2023); https://doi.org/10.1063/5.0148068

[5]     K.-H. Kim, I. Karpov, R. H. Olsson III, D. Jariwala. Wurtzite and fluorite ferroelectric materials for electronic memory, Nature Nanotechnology **18**, 422 (2023); https://doi.org/10.1038/s41565-023-01361-y

[6]     X. Tao, L. Liu, L. Yang and J.-P. Xu, Impacts of HfZrO thickness and anneal temperature on performance of $MoS_2$ negative-capacitance field-effect transistors. Nanotechnology, **32**, 445201 (2021); https://doi.org/10.1088/1361-6528/ac197a

[7]     A. Paul, G. Kumar, A. Das, G. Larrieu, and S. De. Hafnium oxide-based ferroelectric field effect transistors: From materials and reliability to applications in storage-class memory and in-memory computing. J. Appl. Phys. **138**, 010701 (2025); https://doi.org/10.1063/5.0278057

[8]     S.S. Cheema, D. Kwon, N. Shanker, R. Dos Reis, S.-L. Hsu, J. Xiao, H. Zhang, R. Wagner, A. Datar, M. R. McCarter et al. Enhanced ferroelectricity in ultrathin films grown directly on silicon. Nature **580**, 478 (2020); https://doi.org/10.1038/s41586-020-2208-x

[9]     S.S. Cheema, N. Shanker, S.L. Hsu, Y. Rho, C.H. Hsu, V.A. Stoica, Z. Zhang, J.W. Freeland, P. Shafer, C.P. Grigoropoulos, and J. Ciston. Emergent ferroelectricity in subnanometer binary oxide films on silicon. Science, **376**, 648 (2022). https://doi.org/10.1126/science.abm8642

[10]    S.S. Cheema, N. Shanker, L.-C. Wang, C.-H. Hsu, S.-L. Hsu, Y.-H. Liao, M. San Jose, J. Gomez, W. Chakraborty, W. Li et al. Ultrathin ferroic $HfO_2$–$ZrO_2$ superlattice gate stack for advanced transistors. Nature **604**, 65 (2022); https://doi.org/10.1038/s41586-022-04425-6

[11]    S.S. Cheema, N. Shanker, S.-L. Hsu, J. Schaadt, N. M. Ellis, M. Cook, R. Rastogi, R. C.N. Pilawa-Podgurski, J. Ciston, M. Mohamed, S. Salahuddin. Giant energy storage and power density negative capacitance superlattices. Nature **629**, 803 (2024); https://doi.org/10.1038/s41586-024-07365-5

[12]    W. Yang, C. Yu, H. Li, M. Fan, X. Song, H. Ma, Z. Zhou, P. Chang, P. Huang, F. Liu, X. Liu, J. Kang. Ferroelectricity of hafnium oxide-based materials: Current statuses and future prospects from physical mechanisms to device applications. Journal of Semiconductors **44**, 053101 (2023); https://doi.org/10.1088/1674-4926/44/5/053101

[13]    K. P. Kelley, A. N. Morozovska, E. A. Eliseev, Y. Liu, S. S. Fields, S. T. Jaszewski, T. Mimura, J. F. Ihlefeld, S. V. Kalinin. Ferroelectricity in Hafnia Controlled via Surface Electrochemical State. Nature Materials **22**, 1144 (2023); https://doi.org/10.1038/s41563-023-01619-9

[14]    N. Afroze, H. Fahrvandi, G. Ren, P. Kumar, C. Nelson, S. Lombardo, M. Tian, et al. Atomic-scale confinement of strongly charged 180° domain wall pairs in $ZrO_2$. arXiv.2507.18920 (2025); https://doi.org/10.48550/arXiv.2507.18920

[15]    B. Mukherjee, N. S. Fedorova, Jorge Íñiguez-González. Extrinsic nature of the polarization in hafnia ferroelectrics, arXiv.2508.00372 (2025); https://doi.org/10.48550/arXiv.2508.00372

[16]    D.R. Tilley. Finite-size effects on phase transitions in ferroelectrics. In: Ferroelectic Thin Films. ed. C. Paz de Araujo, J.F.Scott and G.W. Teylor.-Amsterdam: Gordon and Breach, 1996.-P.11-45

[17]    G. A. Kourouklis, and E. Liarokapis. Pressure and temperature dependence of the Raman spectra of zirconia and hafnia. Journal of the American Ceramic Society **74**, 520 (1991); https://doi.org/10.1111/j.1151-2916.1991.tb04054.x

[18]    M. Ishigame, and T. Sakurai. Temperature dependence of the Raman spectra of $ZrO_2$. Journal of the American Ceramic Society **60**, 367 (1977); https://doi.org/10.1111/j.1151-2916.1977.tb15561.x

[19]    M. H. Park, Y. H. Lee, H. J. Kim, T. Schenk, W. Lee, K. Do Kim, F. P. G. Fengler, T. Mikolajick, U. Schroeder, and C. S. Hwang. Surface and grain boundary energy as the key enabler of ferroelectricity in nanoscale hafnia-zirconia: a comparison of model and experiment. Nanoscale **9**, 9973 (2017); https://doi.org/10.1039/C7NR02121F

[20]    F. Delodovici, P. Barone, and S. Picozzi. Finite-size effects on ferroelectricity in rhombohedral $HfO_2$. Phys. Rev. B **106**, 115438 (2022); https://doi.org/10.1103/PhysRevB.106.115438

[21] M. H. Park, C.-C. Chung, T. Schenk, C. Richter, K. Opsomer, C. Detavernier, C. Adelmann, J.L. Jones, T. Mikolajick, U. Schroeder. Effect of Annealing Ferroelectric HfO$_2$ Thin Films: In Situ, High Temperature X-Ray Diffraction. Advanced Electronic Materials **4**, 1800091 (2018); https://doi.org/10.1002/aelm.201800091

[22] D. Hyun Lee, Y. Lee, K. Yang, J.Y. Park, S.H. Kim, P.R.S. Reddy, M. Materano, H. Mulaosmanovic, T. Mikolajick, J.L. Jones, U. Schroeder, M.H. Park, Domains and domain dynamics in fluorite-structured ferroelectrics. Appl. Phys. Reviews **8**, 021312 (2021), https://doi.org/10.1063/5.0047977

[23] I. Shlyakhov, K. Iakoubovskii, D. Lin, I. Asselberghs, A. Gaur, G. Delie, and V. Afanas' ev. Energy band alignment in MoS$_2$/HfO$_2$: Transfer-related artifacts and interfacial effects. J. Appl. Phys. **137**, 244304 (2025); https://doi.org/10.1063/5.0279067

[24] T.S. Boscke, S. Teichert, D. Brauhaus, J. Muller, U. Schroder, U. Bottger, and T. Mikolajick, Phase transitions in ferroelectric silicon doped hafnium oxide, Appl. Phys. Lett. **99**, 112904 (2011); https://doi.org/10.1063/1.3636434

[25] Yu Yun, P. Buragohain, M. Li, Z. Ahmadi, Y. Zhang, X. Li, H. Wang, J. Li, P. Lu, L. Tao, H. Wang, Intrinsic ferroelectricity in Y-doped HfO$_2$ thin films, Nature Materials, **21**, 903 (2022); https://doi.org/10.1038/s41563-022-01282-6

[26] W. Yang, C. Yu, H. Li, M. Fan, X. Song, H. Ma, Z. Zhou, P. Chang, P. Huang, F. Liu, X. Liu, J. Kang. Ferroelectricity of hafnium oxide based materials: Current statuses and future prospects from physical mechanisms to device applications. Journal of Semiconductors **44**, 1-45 (2023); https://doi.org/10.1088/1674-4926/44/5/053101

[27] S. Kang, W.-S. Jang, A. N. Morozovska, O. Kwon, Y. Jin, Y.H. Kim, H. Bae, C. Wang, S.H. Yang, A. Belianinov, and S. Randolph, Highly enhanced ferroelectricity in HfO$_2$-based ferroelectric thin film by light ion bombardment. Science, **376**, 731 (2022); https://doi.org/10.1126/science.abk3195

[28] L.-Y. Ma and S. Liu. Structural Polymorphism Kinetics Promoted by Charged Oxygen Vacancies in HfO$_2$. Phys. Rev. Lett. **130**, 096801 (2023); https://doi.org/10.1103/PhysRevLett.130.096801

[29] M.D. Glinchuk, A.N. Morozovska, A. Lukowiak, W. Stręk, M.V. Silibin, D.V. Karpinsky, Y.Kim, and S.V. Kalinin. Possible Electrochemical Origin of Ferroelectricity in HfO$_2$ Thin Films. Journal of Alloys and Compounds, **830**, 153628 (2020); https://doi.org/10.1016/j.jallcom.2019.153628

[30] Oxygen vacancy-induced monoclinic dead layers in ferroelectric HfO$_2$ with metal electrodes. J. Appl. Phys. **137**, 144102 (2025); https://doi.org/10.1063/5.0252663

[31] J. Yoon, Y. Choi and C. Shin. Grain-size adjustment in Hf$_{0.5}$Zr$_{0.5}$O$_2$ ferroelectric film to improve the switching time in Hf$_{0.5}$Zr$_{0.5}$O$_2$-based ferroelectric capacitor. Nanotechnology **35** 135203 (2024); https://doi.org/10.1088/1361-6528/ad0af8

[32] R. Materlik, C. Künneth, and A. Kersch. The origin of ferroelectricity in Hf$_{1-x}$Zr$_x$O$_2$: A computational investigation and a surface energy model, J. Appl. Phys. **117**, 134109 (2015); http://dx.doi.org/10.1063/1.4916707

[33] E. A. Eliseev, I. V. Kondakova, Y. O. Zagorodniy, H. V. Shevliakova, O. V. Leshchenko, V. N. Pavlikov, M. V. Karpets, L. P. Yurchenko, and A. N. Morozovska, The origin of the ferroelectric-like

orthorhombic phase in oxygen-deficient HfO2-y nanoparticles. Semiconductor Physics, Optoelectronics and Quantum Electronics, **28** (2), 134-144 (2025), https://doi.org/10.15407/spqeo28.02.134

[34] Y. Qi, K. M. Rabe, Competing phases of $HfO_2$ from multiple unstable flat phonon bands of an unconventional high-symmetry phase (2024); https://doi.org/10.48550/arXiv.2412.16792

[35] Y. Qi, S. Singh, and K. M. Rabe. Polarization switching in ferroelectric $HfO_2$ from first-principles lattice mode analysis. Phys. Rev. B, **111**, 134106 (2025); https://doi.org/10.1103/PhysRevB.111.134106

[36] E. A. Eliseev, S. V. Kalinin, A. N. Morozovska. Ferro-ionic States and Domains Morphology in $Hf_xZr_{1-x}O_2$ Nanoparticles. Journal of Applied Physics, **137 (3),** 034103 (2025); https://doi.org/10.1063/5.0243067

[37] K. Fujimoto, Y. Sato, Y. Fuchikami, R. Teranishi, and Kenji Kaneko. Orthorhombic-like atomic arrangement in hafnium-oxide-based nanoparticles. Journal of the American Ceramic Society **105**, 2823 (2022); https://doi.org/10.1111/jace.18242

[38] E. A. Eliseev, Y. O. Zagorodniy, V. N. Pavlikov, O. V. Leshchenko, H. V. Shevilakova, M. V. Karpec, A. D. Yaremkevych, O. M. Fesenko, S. V. Kalinin, and A. N. Morozovska. Phase diagrams and polarization reversal in nanosized $Hf_xZr_{1-x}O_{2-y}$, AIP Advances, **14**, 055224 (2024); https://doi.org/10.1063/5.0209123

[39] O. S. Pylypchuk, I. V. Fesych, V. V. Vainberg, Y. O. Zagorodniy, V. I. Styopkin, J. M. Gudenko, I. V. Kondakova, L. P. Yurchenko, V. N. Pavlikov, A. O. Diachenko, M. M. Koptiev, M. D. Volnyanskii, V. V. Laguta, E. A. Eliseev, M. P. Trubitsyn, and A. N. Morozovska. Resistive switching and charge accumulation in $Hf_{0.5}Zr_{0.5}O_2$ nanoparticles. Journal of Physical Chemistry **29**, 31, 14299 (2025); https://doi.org/10.1021/acs.jpcc.5c04140

[40] W. Heywang. Semiconducting Barium Titanate. J. Materials Science **6**, 1214 (1971); https://doi.org/10.1007/BF00550094

[41] See Supplementary Materials for details [URL will be provided by Publisher]

[42] S. Kang, W.-S. Jang, A. N. Morozovska, O. Kwon, Y. Jin, Y.H. Kim, H. Bae, C. Wang, S.H. Yang, A. Belianinov, and S. Randolph, Highly enhanced ferroelectricity in $HfO_2$-based ferroelectric thin film by light ion bombardment. Science, **376**, 731 (2022); https://doi.org/10.1126/science.abk3195

[43] K. P. Kelley, A. N. Morozovska, E. A. Eliseev, Y. Liu, S. S. Fields, S. T. Jaszewski, T. Mimura, J. F. Ihlefeld, S. V. Kalinin. Ferroelectricity in Hafnia Controlled via Surface Electrochemical State. Nature Materials **22**, 1144 (2023); https://doi.org/10.1038/s41563-023-01619-9

[44] L.-Y. Ma and S. Liu. Structural Polymorphism Kinetics Promoted by Charged Oxygen Vacancies in $HfO_2$. Phys. Rev. Lett. **130**, 096801 (2023); https://doi.org/10.1103/PhysRevLett.130.096801

[45] A. N. Morozovska, A. V. Bodnaruk, O. S. Pylypchuk, D. O. Stetsenko, A. D. Yaremkevich, O. V. Leshchenko, V. N. Pavlikov, Y. O. Zagorodniy, V. V. Laguta, L. P. Yurchenko, L. Demchenko, M. V. Karpets, O. M. Fesenko, V. V. Vainberg, and E. A. Eliseev, Multiferroicity of oxygen-deficient $Hf_xZr_{1-x}O_{2-y}$ nanoparticles (2025); https://doi.org/10.48550/arXiv.2509.21621

[46]     S. Wang, Li X., Jia Y., et al.. Direct observation of charge density and electronic polarization in fluorite ferroelectrics by 4D-STEM. The Innovation Materials **2**, 100068, (2024); https://doi.org/10.59717/j.xinn-mater.2024.100068

[47]     C. Yu, H. Ma, M. Li, F. Liu, X. Ding, Y. Zhao, H. Li et al. Insights into the origin of robust ferroelectricity in $HfO_2$-based thin films from the order-disorder transition driven by vacancies. Phys. Rev. Applied **22**, 024028 (2024); https://doi.org/10.1103/PhysRevApplied.22.024028

[48]     B. Zeng, L. Yin, R. Liu, C. Ju, Q. Zhang, Zhibin Yang, Shuaizhi Zheng et al. Multiple Polarization States in $Hf_{1-x}Zr_xO_2$ Thin Films by Ferroelectric and Antiferroelectric Coupling. Advanced Materials **37**, 2411463 (2025); https://doi/10.1002/adma.202411463

[49]     A. N. Morozovska, O. S. Pylypchuk, S. Ivanchenko, E. A. Eliseev, H. V. Shevliakova, L. M. Korolevich, L. P. Yurchenko, O. V. Shyrokov, N. V. Morozovsky, V. N. Poroshin, Z. Kutnjak, and V. V. Vainberg. Size-induced High Electrocaloric Response of the Dense Ferroelectric Nanocomposites. Ceramics International **50**, 11743 (2024); https://doi.org/10.1016/j.ceramint.2024.01.079

[50]     A.N. Morozovska, M.D. Glinchuk, E.A. Eliseev. Phase transitions induced by confinement of ferroic nanoparticles. Phys. Rev. B **76**, 014102 (2007); https://doi.org/10.1103/PhysRevB.76.014102

[51]     E. A. Eliseev, A. N. Morozovska, S. V. Kalinin, and D. R. Evans. Strain-Induced Polarization Enhancement in $BaTiO_3$ Core-Shell Nanoparticles. Phys. Rev. **B**. **109**, 014104 (2024); https://doi.org/10.1103/PhysRevB.109.014104

[52]     J. Robertson. High dielectric constant oxides. Eur. Phys. J. Appl. Phys. **28**, 265 (2004); https://doi.org/10.1051/epjap:2004206

[53]     B. Saini, F. Huang, Y.-Y. Choi, Z. Yu, V. Thampy, J.D. Baniecki, W. Tsai, and P.C. McIntyre. Field-Induced Ferroelectric Phase Evolution During Polarization "Wake-Up" in $Hf_{0.5}Zr_{0.5}O_2$ Thin Film Capacitors. Advanced Electronic Materials **9**, 2300016 (2023); https://doi.org/10.1002/aelm.202300016

[54]     S. Yoneda, T. Hosokura, T. Usui, M. Kimura, A. Ando, and K. Shiratsuyu. High dielectric permittivity of HfO2-based films with (La,Bi,Nb) substitution. Japanese Journal of Applied Physics **57**, 11UF03 (2018); https://doi.org/10.7567/JJAP.57.11UF03

[55]     K. Ishikawa, K. Yoshikawa, and N. Okada. Size effect on the ferroelectric phase transition in $PbTiO_3$ ultrafine particles. Phys. Rev. B **37**, 5852 (1988); https://doi.org/10.1103/PhysRevB.37.5852

[56]     B. Qu, J. Bin, W. Yuguo, Z. Peilin, and Z. Weilie. Size and temperature dependence of dielectric constant of ultrafine PbTiO3 particles. Chinese Physics Letters, **11**, 514 (1994). https://doi.org/10.1088/0256-307X/11/8/014

[57]     E. Erdem, H.-C. Semmelhack, R. Böttcher, H. Rumpf, J. Banys, A. Matthes, H.-J. Gläsel, D. Hirsch, and E. Hartmann. Study of the tetragonal-to-cubic phase transition in $PbTiO_3$ nanopowders. J. Phys.: Condens. Matter **18,** 3861 (2006); https://doi.org/10.1088/0953-8984/18/15/028

[58]     G. Arlt, D. Hennings, G. de With, Dielectric properties of fine-grained barium titanate ceramics. J. Appl. Phys. S8, 1619 (1985); https://doi.org/10.1063/1.336051

[59] Y. Tan, J. Zhang, Y. Wu, C. Wang, V. Koval, B. Shi, H. Ye, R. McKinnon, G. Viola, and H. Yan. Unfolding grain size effects in barium titanate ferroelectric ceramics. *Scientific reports* **5**, 9953 (2015); https://doi.org/10.1038/srep09953

[60] T. Hoshina, Size effect of barium titanate: fine particles and ceramics. Journal of the ceramic society of Japan **121**, 156 (2013); https://doi.org/10.2109/jcersj2.121.156

[61] J. Petzelt, I. Rychetsky, D. Nuzhnyy, Dynamic ferroelectric–like softening due to the conduction in disordered and inhomogeneous systems: giant permittivity phenomena. Ferroelectrics, **426**, 171-193 (2012); https://doi.org/10.1080/00150193.2012.671732

[62] J. Petzelt, D. Nuzhnyy, V. Bovtun, M. Savinov, M. Kempa, I. Rychetsky, Broadband dielectric and conductivity spectroscopy of inhomogeneous and composite conductors. Phys. Stat. Sol. A **210**, 2259-2271 (2013); https://doi.org/10.1002/pssa.201329288

[63] O. S. Pylypchuk, S. E. Ivanchenko, M. Y. Yelisieiev, A. S. Nikolenko, V. I. Styopkin, B. Pokhylko, V. Kushnir, D. O. Stetsenko, O. Bereznykov, O. V. Leschenko, E. A. Eliseev, V. N. Poroshin, N. V. Morozovsky, V. V. Vainberg, and A. N. Morozovska. Behavior of the Dielectric and Pyroelectric Responses of Ferroelectric Fine-Grained Ceramics. Journal of the American Ceramic Society, **108**, e20391 (2025); https://doi.org/10.1111/jace.20391

[64] H. Han, Ch. Voisin, S. Guillemet-Fritsch, P. Dufour, Ch. Tenailleau, Ch. Turner, and J.C. Nino. Origin of colossal permittivity in $BaTiO_3$ via broadband dielectric spectroscopy. J. Appl. Phys. **113**, 024102 (2013); https://doi.org/10.1063/1.4774099

[65] L. Liu, S. Ren, J. Liu, F. Han, J. Zhang, B. Peng, D. Wang, A. A. Bokov, and Z.-G. Ye, Localized polarons and conductive charge carriers: Understanding $CaCu_3Ti_4O_{12}$ over a broad temperature rangePhys. Rev. B **99**, 094110 (2019), https://doi.org/10.1103/PhysRevB.99.094110

[66] K.W. Wagner, Erklärung der dielektrischen Nachwirkungsvorgänge auf Grund Maxwellscher Vorstellungen. Arch Elektrotech. **2**, 371 (1914); https://doi.org/10.1007/BF01657322

[67] L. E. Cross, Relaxor ferroelectrics. Ferroelectrics **76**, 241 (1987); https://doi.org/10.1080/00150198708016945

[68] A. A. Bokov and Z.-G. Ye, Low-frequency dielectric spectroscopy of the relaxor ferroelectric $Pb(Mg_{1/3}Nb_{2/3})O_3$-$PbTiO_3$, Phys. Rev. B **65**, 144112 (2002); https://doi.org/10.1103/PhysRevB.65.144112

[69] E. Courtens, Vogel-Fulcher scaling of the susceptibility in a mixed-crystal proton glass. Physical review letters **52**, 69 (1984); https://doi.org/10.1103/PhysRevLett.52.69

[70] E. Courtens, Scaling dielectric data on $Rb_{1-x}(NH_4)_x H_2PO_4$ structural glasses and their deuterated isomorphs. Phys. Rev. B **33**, 2975 (1986); https://doi.org/10.1103/physrevb.33.2975

[71] D. Viehland, M. Wuttig and L. E. Cross. The glassy behavior of relaxor ferroelectrics, Ferroelectrics, **120**, 71-77(1991); http://doi.org/10.1080/00150199108216802

[72] I. Rivera, Ashok Kumar, N. Ortega, R. S. Katiyar, and S, Lushnikov. "Divide line between relaxor, diffused ferroelectric, ferroelectric and dielectric." Solid State Communications **149**, 172-176 (2009); https://doi.org/10.1016/j.ssc.2008.10.026

[73] S.L. Bravina et.al., Study of peculiarities of pyroelectric properties of prustite and pyrargyrite crystals by the pyroelectric luminescence method. Ukrainian Journal of Physics, **32** (10), 1964 (1987)

[74] S. B Lang, Pyroelectricity: from ancient curiosity to modern imaging tool. Physics today 58, 31 (2005); https://doi.org/10.1063/1.2062916

[75] I.P. Zvyagin. In: *Charge Transport in Disordered Solids with Applications in Electronics*, ed. by S. Baranovski (John Wiley & Sons, Chichester, 2006) p. 339.

[76] M. Pollak, T.H. Geballe. Low-Frequency Conductivity Due to Hopping Processes in Silicon. Phys. Rev. **122**, 1742 (1961); https://doi.org/10.1103/PhysRev.122.1742

[77] N. F. Mott and W. D. Twose, The Theory of Impurity Conduction, Advances Phys. **10** (38), 107 (1961); https://doi.org/10.1080/00018736100101271

[78] T. Kolodiazhnyi, A. Petric, M. Niewczas, C. Bridges, A. Safa-Sefat and J. E. Greedan. Thermoelectric power, Hall effect, and mobility of *n*-type BaTiO3. Phys. Rev. B **68**, 085205 (2003); https://doi.org/10.1103/PhysRevB.68.085205

[79] M. Maglione. Polarons, free charge localisation and effective dielectric permittivity in oxides. V.S. Vikhnin and G. K. Liu. Springer Series of Topics in Solid-State Sciences, Springer Verlag, 2010, Topics in Solid-State Sciences. hal-00493298

[80] N. F. Mott, Conduction in non-crystalline materials. Philosophical Magazine. **19** (160), 835 (1969); https://doi.org/10.1080/14786436908216338

[81] T.C. Choy. Effective Medium Theory. Oxford (UK): Clarendon Press; 1999. ISBN 978-0-19-851892-1.

[82] L.D. Landau, L.P. Pitaevskii, E.M. Lifshitz. Course of Theoretical Physics. 2$^{nd}$ ed. Vol. 8, Electrodynamics of continuous media. Translated by J.S. Bell, M.J. Kearsley, J.B. Sykes. Elsevier: Oxford; 2013.

[83] J.C.M. Garnett. Colours in metal glasses and in metallic films. Philos Trans R. Soc. Ser. A. **203**, 385 (1904); https://doi.org/10.1098/rsta.1904.0024.

[84] D.A.G. Bruggeman. Berechnung verschiedener physikalischer Konstanten von heterogenen Substanzen. I. Dielektrizitätskonstanten und Leitfähigkeiten der Mischkörper aus isotropen Substanzen, Ann Phys. **416**, 636 (1935); https://doi.org/10.1002/andp.19354160705

[85] R. Simpkin. Derivation of Lichtenecker's logarithmic mixture formula from Maxwell's equations. IEEE Transactions on Microwave Theory and Techniques. **58**, 545 (2010); https://doi.org/10.1109/TMTT.2010.2040406.

[86] G.L. Carr, S. Perkowitz, D.B. Tanner. Far-infrared properties of inhomogeneous materials. In: Button KJ, editor. Infrared and millimeter waves, Orlando: Academic Press, **13**, 171-263 (1985)

[87] O. S. Pylypchuk, S. E. Ivanchenko, Y. O. Zagorodniy, M. E. Yelisieiev, O. V. Shyrokov, O.V. Leschenko, O. Bereznykov, D. Stetsenko, S. D. Skapin, E. A. Eliseev, V. N. Poroshin, V. V. Vainberg, and A. N. Morozovska. Relaxor-like Behavior of the Dielectric Response of Dense Ferroelectric Composites. Ceramics International **50,** Issue 22, Part A, 45465-45478 (2024); https://doi.org/10.1016/j.ceramint.2024.08.385

[88]   J. Petzelt, D. Nuzhnyy, V. Bovtun, et al. Origin of the colossal permittivity of (Nb+ In) co-doped rutile ceramics by wide-range dielectric spectroscopy. Phase Transitions. **91** (9-10), 932 (2018); https://doi.org/10.1080/01411594.2018.1501801.

[89]   I. Rychetský, D. Nuzhnyy, J. Petzelt. Giant permittivity effects from the core–shell structure modeling of the dielectric spectra. Ferroelectrics. **9** (1), 9-20 (2020); https://doi.org/10.1080/00150193.2020.1791659.

[90]   A. N. Morozovska, E. A. Eliseev, A. Biswas, N. V. Morozovsky, and S. V. Kalinin. Effect of surface ionic screening on polarization reversal and phase diagrams in thin antiferroelectric films for information and energy storage. Phys. Rev. Applied **16**, 044053 (2021); http://doi.org/10.1103/PhysRevApplied.16.044053

[91]   A. Biswas, A. N. Morozovska, M. Ziatdinov, E. A. Eliseev and S. V. Kalinin. Multi-objective Bayesian optimization of ferroelectric materials with interfacial control for memory and energy storage applications. J. Appl. Phys. **130**, 204102 (2021); https://doi.org/10.1063/5.0068903

[92]   R. Blinc, B. Zeks, Soft Mode in Ferroelectrics and Antiferroelectrics; North-Holland Publishing Company, Amsterdam, Oxford, 1974.

[93]   A. K. Tagantsev, K. Vaideeswaran, S. B. Vakhrushev, A. V. Filimonov, R. G. Burkovsky, A. Shaganov, D. Andronikova, A. I. Rudskoy, A. Q. R. Baron, H. Uchiyama, et al. The Origin of Antiferroelectricity in PbZrO3. Nature Commun. **4**, article number 2229 (2013); https://doi.org/10.1038/ncomms3229

[94]   K.M. Rabe, Antiferroelectricity in Oxides: A Reexamination. In Functional Metal Oxides; John Wiley & Sons, Ltd; pp 221–244 (2013); https://doi.org/10.1002/9783527654864.ch7

[95]   J. Hlinka, T. Ostapchuk, E. Buixaderas, C. Kadlec, P. Kuzel, I. Gregora, J. Kroupa, M. Savinov, A. Klic, J. Drahokoupil, et al. Multiple Soft-Mode Vibrations of Lead Zirconate. Phys. Rev. Lett. **112**, 197601 (2014); https://doi.org/10.1103/PhysRevLett.112.197601

[96]   A. N. Morozovska, M. V. Strikha, K. P. Kelley, S. V. Kalinin, and E. A. Eliseev. Effective Landau-type model of a $Hf_xZr_{1-x}O_2$-graphene nanostructure, Phys. Rev. Appl. **20**, 054007 (2023); https://doi.org/10.1103/PhysRevApplied.20.054007

[97]   A. N. Morozovska, E. A. Eliseev, Yu. M. Vysochanskii, V. V. Khist, and D. R. Evans. Screening-Induced Phase Transitions in Core-Shell Ferroic Nanoparticles. Phys.Rev. Materials **6**, 124411 (2022); https://doi.org/10.1103/PhysRevMaterials.6.124411

[98]   J. H. Barrett, Dielectric Constant in Perovskite Type Crystals, Phys. Rev. **86**, 118 (1952); https://doi.org/10.1103/PhysRev.86.118

[99]   K. Koepernik and H. Eschrig, Full-potential nonorthogonal local-orbital minimum-basis band-structure scheme, Phys. Rev. B **59**, 1743 (1999); https://doi.org/10.1103/PhysRevB.59.1743

[100]   J. Müller, T. S. Böscke, U. Schröder, S. Mueller, D. Bräuhaus, U. Böttger, L. Frey, and T. Mikolajick, Ferroelectricity in simple binary $ZrO_2$ and $HfO_2$, Nano Lett. **12** (8), 4318 (2012); https://doi.org/10.1021/nl302049k

[101] E. H. Kisi and C. J. Howard, Crystal structures of zirconia phases and their inter-relation, Key Eng. Mater. **153–154**, 1–36 (1998); https://doi.org/10.4028/www.scientific.net/kem.153-154.1

[102] E. H. Kisi, C. J. Howard, and R. J. Hill, Crystal structure of orthorhombic zirconia in partially stabilized zirconia, J. Am. Ceram. Soc. **72** (9), 1757 (1989); https://doi.org/10.1111/j.1151-2916.1989.tb06322.x

[103] D. A. Freedman, D. Roundy, and T. A. Arias. Elastic effects of vacancies in strontium titanate: Short- and long-range strain fields, elastic dipole tensors, and chemical strain. Phys. Rev. B **80**, 064108 (2009); https://doi.org/10.1103/PhysRevB.80.064108

[104] Y. Kim, A. S. Disa, T. E. Babakol, and J. D. Brock. Strain screening by mobile oxygen vacancies in $SrTiO_3$. Appl. Phys. Lett. **96**, 251901 (2010); https://doi.org/10.1063/1.3455157

[105] V.A. Shchukin and D. Bimberg. Spontaneous ordering of nanostructures on crystal surfaces. Reviews of Modern Physics, **71**, 1125 (1999); https://doi.org/10.1103/RevModPhys.71.1125

[106] W. Ma. Surface tension and Curie temperature in ferroelectric nanowires and nanodots. Appl. Phys. A **96**, 915 (2009); http://dx.doi.org/10.1007/s00339-009-5246-7

[107] W. Ma. Surface tension and Curie temperature in ferroelectric nanowires and nanodots (Erratum). Appl. Phys. A **96**, 1035 (2009); https://doi.org/10.1007/s00339-009-5316-x

[108] A. K. Tagantsev, G. Gerra, and N. Setter, Short-range and long-range contributions to the size effect in metal-ferroelectric-metal heterostructures, Phys. Rev. B **77**, 174111 (2008); https://doi.org/10.1103/PhysRevB.77.174111

[109] W. Zhu, J. Hayden, F. He, J.-I. Yang; P. Tipsawat, M. D. Hossain, J.-P. Maria, and S. Trolier-McKinstry. Strongly temperature dependent ferroelectric switching in AlN, $Al_{1-x}Sc_xN$, and $Al_{1-x}B_xN$ thin films. Applied Physics Letters **119**, 062901 (2021); https://doi.org//10.1063/5.0057869

[110] N. Wolff, G. Schönweger, I. Streicher, M.R. Islam, N. Braun, P. Straňák, L. Kirste, M. Prescher, A. Lotnyk, H. Kohlstedt, and S. Leone, Demonstration and STEM Analysis of Ferroelectric Switching in MOCVD-Grown Single Crystalline $Al_{0.85}Sc_{0.15}N$. Advanced Physics Research **3**, 2300113 (2024), https://doi.org//10.1002/apxr.202300113

[111] J. Yang, A. V. Ievlev, A. N. Morozovska, E. Eliseev, J. D Poplawsky, D. Goodling, R. J. Spurling, J.-P. Maria, S. V. Kalinin, Y. Liu. Coexistence and interplay of two ferroelectric mechanisms in $Zn_{1-x}Mg_xO$. Advanced Materials **36** (39), 2404925 (2024), https://doi.org/10.1002/adma.202404925

[112] C. H. Skidmore, R. J. Spurling, J. Hayden, S. M. Baksa, D. Behrendt, D. Goodling, J. L. Nordlander, A. Suceava, J. Casamento, B. Akkopru-Akgun, S. Calderon, I. Dabo, V. Gopalan, K. P. Kelley, A. M. Rappe, S. Trolier-McKinstry, E. C. Dickey & J.-P. Maria. Proximity ferroelectricity in wurtzite heterostructures, Nature, (2024), https://doi.org/10.1038/s41586-024-08295-y

[113] E. A. Eliseev, A. N. Morozovska, J.-P. Maria, L.-Q. Chen, and V. Gopalan. A Thermodynamic Theory of Proximity Ferroelectricity. Physical Review X, **15**, 021058 (2025), https://doi.org/10.1103/PhysRevX.15.021058

[114] Y. Jiang, E. Parsonnet, A. Qualls, W. Zhao, S. Susarla, D. Pesquera, A. Dasgupta, M. Acharya, H. Zhang, T. Gosavi, C.C. Lin, D. E. Nikonov, H. Li, I. A. Young, R. Ramesh and L. W. Martin Enabling ultra-

low-voltage switching in BaTiO$_3$. Nature materials, **21**, 779 (2022); https://doi.org/10.1038/s41563-022-01266-6

[115] https://www.wolfram.com/mathematica, for the codes see https://notebookarchive.org/2025-08-7fidfp8

[116] W. Heywang. Semiconducting Barium Titanate. J. Materials Science **6**, 1214 (1971); https://doi.org/10.1007/BF00550094